\documentclass[10pt,journal,compsoc]{IEEEtran}
\usepackage{xurl}
\usepackage{xcolor}
\usepackage{hyperref}
\usepackage{enumitem}
\usepackage[linesnumbered, ruled,vlined]{algorithm2e}
\usepackage{etoolbox}
\usepackage{makecell}

\makeatletter
\patchcmd{\@algocf@start}
  {-1.5em}
  {0pt}
  {}{}
\makeatother

\ifCLASSOPTIONcompsoc
  \usepackage[nocompress]{cite}
\else
  \usepackage{cite}
\fi
\ifCLASSINFOpdf
  \usepackage[pdftex]{graphicx}
\else
\fi
\begin{document}
%
\title{D-LNBot: A Scalable, Cost-Free and Covert\\ Hybrid Botnet on Bitcoin's Lightning Network*}
%
%
%
%

\author{Ahmet~Kurt,
        Enes~Erdin,
        Kemal~Akkaya,
        A.~Selcuk~Uluagac,
        and~Mumin~Cebe~
\IEEEcompsocitemizethanks{\IEEEcompsocthanksitem A. Kurt, K. Akkaya, and A. S. Uluagac are with the Department of Electrical and Computer Engineering, Florida International University, Miami, FL, 33174.
E-mail: \{akurt005, kakkaya, suluagac\}@fiu.edu
\IEEEcompsocthanksitem E. Erdin is with the Department of Computer Science, University of Central Arkansas. E-mail: eerdin@uca.edu
\IEEEcompsocthanksitem M. Cebe is with the Department of Computer Science, Marquette University. E-mail: mumin.cebe@marquette.edu
\IEEEcompsocthanksitem *A preliminary version \cite{kurt2020lnbot} of this work was published in the proceedings of the 25th European Symposium on Research in Computer Security (ESORICS 2020).}

}

%
%

\markboth{}%
{Shell \MakeLowercase{\textit{Kurt et al.}}: D-LNBot: A Scalable, Cost-Free Covert Hybrid Botnet on Bitcoin's Lightning Network}
%



\IEEEtitleabstractindextext{%
\begin{abstract}
While various covert botnets were proposed in the past, they still lack complete anonymization for their servers/botmasters or suffer from slow communication between the botmaster and the bots. In this paper, we first propose a new generation hybrid botnet that covertly and efficiently communicates over Bitcoin Lightning Network (LN), called LNBot. Exploiting various anonymity features of LN, we show the feasibility of a scalable two-layer botnet which completely anonymizes the identity of the botmaster. In the first layer, the botmaster anonymously sends the commands to the command and control (C\&C) servers through regular LN payments. Specifically, LNBot allows botmaster's commands to be sent in the form of surreptitious multi-hop LN payments, where the commands are either encoded with the payments or attached to the payments to provide covert communications. In the second layer, C\&C servers further relay those commands to the bots in their mini-botnets to launch any type of attacks to victim machines. We further improve on this design by introducing D-LNBot; a distributed version of LNBot that generates its C\&C servers by infecting users on the Internet and forms the C\&C connections by opening channels to the existing nodes on LN. In contrary to the LNBot, the whole botnet formation phase is distributed and the botmaster is never involved in the process. By utilizing Bitcoin's Testnet and the new message attachment feature of LN, we show that D-LNBot can be run for free and commands are propagated faster to all the C\&C servers compared to LNBot. We presented proof-of-concept implementations for both LNBot and D-LNBot on the actual LN and extensively analyzed their delay and cost performance. Finally, we also provide and discuss a list of potential countermeasures to detect LNBot and D-LNBot activities and minimize their impacts. 
\end{abstract}

}

\maketitle

\IEEEdisplaynontitleabstractindextext

%
\IEEEpeerreviewmaketitle

\IEEEraisesectionheading{\section{Introduction}\label{sec:introduction}}

%
%
%
%
\IEEEPARstart{B}{otnets} are networks of computing devices infected with malicious software that is under the control of an attacker, known as bot herder or \textit{botmaster} \cite{silva2013botnets}. The owner of the botnet controls the \textit{bots} (i.e., devices that become part of the botnet) through command and control \textit{(C\&C)} server(s) which can communicate with the bots using a C\&C channel and can launch various attacks through these bots, including, but not limited to, denial of service (DoS) attacks, information and identity theft, sending spam messages, and other activities. Naturally, a botmaster's goal is to make it difficult for law enforcement to detect and prevent malicious operations. Therefore, establishing a secure C\&C infrastructure and hiding the identities of the C\&C servers play a key role in the long-lasting operation of botnets. 

Numerous botnets have been proposed and deployed in the past \cite{wang2008advanced, soltan2018blackiot}. Regardless of their communication infrastructure being centralized or peer-to-peer, existing botnet C\&C channels and servers have the challenge of remaining hidden and being resistant against legal authorities' actions. Such problems motivate hackers to always explore more promising venues for finding new C\&C channels with the ever-increasing number of communication options on the Internet. One such platform is the environment where cryptocurrencies, such as Bitcoin \cite{nakamoto2008bitcoin}, is exchanged. As Bitcoin offers some degree of anonymity, exploiting it as a C\&C communication channel has already been tried for creating new botnets~\cite{ali2015zombiecoin,frkat2018chainchannels}. While some of these Bitcoin-based botnets addressed the long transaction validation times, they still announce the commands publicly, where the botnet activity can be traced by any observer with the help of the Bitcoin addresses or nonce values of the transactions. By using Bitcoin for botnet C\&C communication, the history of malicious activities are recorded on the blockchain forever.

Nonetheless, the issues regarding the public announcement of commands and leaving traces in the blockchain are already addressed with a Bitcoin payment channel network called Lightning Network (LN) \cite{poon2016bitcoin}. LN enables off-chain transactions (i.e., transactions which are not announced and thus not recorded on the blockchain) in order to speed up the transaction by eliminating the confirmation time and decreasing fees associated with that transaction. Additionally, users' identities are more anonymous since the transactions are not announced publicly. In this paper, we show that LN can be exploited as an ideal C\&C infrastructure for botnets with all the aforementioned features (i.e., faster transactions, decreased costs, more anonymity). Specifically, LN offers botmasters numerous advantages over existing techniques: First, LN provides very high anonymity since the transactions on the off-chain channels are not recorded on the blockchain. Thus, a botmaster can covertly communicate with the C\&C server(s). Second, the revelation of a server does not reveal other servers, and an observer cannot enumerate the size of the botnet. Most importantly, C\&C communication over the LN cannot be censored. 

Although LN is a fast-growing emerging payment network, it only has around 16,000 nodes which may not be ideal for large-scale botnets. Therefore, we propose a \textit{two-layer hybrid} botnet to use LN as an infrastructure to maintain a network of C\&C servers each of which can run its own botnet. The use of multiple C\&C servers has been around for a while \cite{zeidanloo2009botnet}. However, the communication with these servers was still assumed to be through the existing communication infrastructures which impairs the servers' anonymity. Therefore, LN can be utilized to further strengthen the anonymity.

Hence, this paper first presents \textit{LNBot}, which is the first botnet that utilizes LN infrastructure for its communications between the botmaster and C\&C servers with a two-layer hybrid architecture. Specifically, at the first layer, a botmaster maintains multiple C\&C servers, which are nodes on the LN that have specialized software to control the bots under them. Essentially, each C\&C server controls an independent isolated \textit{mini-botnet} at the second layer. These mini-botnets are controlled using a specific C\&C infrastructure that can rely on traditional means such as stenography, IRC channel, DNS, social media, etc. Botmaster sends the commands to the C\&C servers covertly through LN. This two-layer command and control mechanism not only enables scalability, but also minimizes the burden on each C\&C server, which increases their anonymity. Second, we present \textit{iLNBot} which stands for \textit{improved version of LNBot} that has significantly reduced cost and latency overheads. To achieve this, we employ a new functionality in LN that enables sending messages to other LN nodes by attaching them to the regular LN payments. 

However, the proposed LNBot and iLNBot still assume a centralized communication model between the botmaster and C\&C servers. To further strengthen the anonymity, we show that a distributed version, namely \textit{D-LNBot} can be created which forms itself over existing LN nodes. Specifically, we propose to utilize a C\&C discovery mechanism where the new C\&C servers that join the network can discover some of the previously joined C\&C servers and vice versa. This is achieved by using LN channels that are opened to public LN nodes following a specific rule which is only recognizable by other C\&C servers. In this way, the management burden on the botmaster is almost completely removed. Additionally, proposed LNBot utilizes LN on Bitcoin Mainnet which incurs non-negligible costs to the botmaster for sending the commands. In D-LNBot, we show that in fact, Bitcoin's Testnet can also be used instead of the Mainnet to remove any costs that were present in LNBot. Since Testnet Bitcoins can be obtained for free, botmaster will not have to spend any money to send commands nor fund C\&C servers' channels. All these versions follow the same hybrid architecture where the C\&C servers control their own mini-botnet through various C\&C infrastructures.

To demonstrate the feasibility of the concept, we implemented the all three versions (i.e., LNBot, iLNBot, D-LNBot) using real LN nodes on Bitcoin's Testnet. For LNBot, we utilize a one-to-many architecture (i.e., botmaster sends the commands to all C\&C servers separately) for sending the commands. We show that by encoding the commands into payments sent over LN, one can successfully send commands to the C\&C servers that are part of the LN. The C\&C servers further relay those commands to the bots they control to launch any type of attack to victim machines. For iLNBot, we use the same one-to-many architecture but the commands are embedded into payments rather than getting encoded. This change let us achieve a 98\% reduction on the command sending cost and 99\% reduction on the command sending delay. Finally for D-LNBot, we show that the commands can be sent to all the C\&C servers with a single payment without spending any money. On top of sending the commands for free, D-LNBot propagates the commands to the C\&C servers faster than LNBot and iLNBot.

Our contributions in this work are as follows:
\begin{itemize}[leftmargin=*]
    \item We first propose LNBot which is a covert and hybrid botnet that uses Bitcoin's LN for its C\&C communications. LN's strong anonymity features makes it very challenging to stop the botnet.
    \item Next, we propose iLNBot which is a significantly improved version of LNBot in terms of cost and time spent on sending the commands to the bots.
    \item We then propose D-LNBot which is a distributed version of LNBot that is superior in cost, delay and resiliency aspects. Specifically, D-LNBot forms itself over LN with minimum botmaster intervention, sends the commands to the bots for free, spends significantly less time to propagate the commands to all the bots.
    \item We present proof of concept implementations for all three versions and extensively analyze their performance metrics. Some implementation details can be found in our GitHub repository at: \url{https://github.com/startimeahmet/D-LNBot}.
    \item Finally, all these features of LNBot and D-LNBot make them botnets that need to be taken seriously therefore we provide a list of countermeasures that may help detect LNBot and D-LNBot activities and minimize damages from them.
\end{itemize}

The rest of the paper is organized as follows: Related work is given in Section \ref{sec:relatedwork}. In Section \ref{sec:background}, we give some background information about LN. Section \ref{sec:threatmodel} describes the threat model. In Section \ref{sec:LNBot}, we describe the architecture and construction of our proposed LNBot. Similarly, in Section \ref{sec:D-LNBot}, we describe the details of our proposed D-LNBot. Section \ref{sec:POC} is dedicated to the proof-of-concept implementations of LNBot and D-LNBot in real world settings while Section \ref{sec:evaluation} presents their evaluation results. In Section \ref{sec:countermeasures}, possible countermeasures for LNBot and D-LNBot are discussed. Finally, we conclude the paper in Section \ref{sec:conclusion}.

\section{Related Work}
\label{sec:relatedwork}
Botnets have been around for a long time and there have been even surveys classifying them \cite{silva2013botnets, bailey2009survey}. While early botnets used IRC, later botnets focused on P2P C\&C channels for resiliency \cite{grizzard2007peer}. Our proposed LNBot and D-LNBot fall under covert botnets which became popular much later. As an example, Nagaraja et al. proposed Stegobot, a covert botnet using social media networks as a command and control channel \cite{nagaraja2011stegobot}. Pantic et al. proposed a covert botnet command and control using Twitter \cite{pantic2015covert}. Tsiatsikas et al. proposed SDP-based covert channel for botnet communication \cite{tsiatsikas2015hidden}. Calhoun et al. presented a MAC layer covert channel based on WiFi \cite{calhoun2012802}. A recent work by Wu et al. \cite{wu2021simple} presents a mobile covert botnet network. Another covert botnet design based on social networks was recently proposed by Wang et al. \cite{wang2022deepc2}. Tian et al. proposed DLchain \cite{tian2020dlchain}, a covert channel utilizing the Bitcoin network.

Recently, there have been proposals on using the Bitcoin blockchain for botnet C\&C communication \cite{bock2019assessing}. For instance, Roffel et al.~\cite{Roffel} came up with the idea of controlling a computer worm using the Bitcoin blockchain. Another work\cite{sweeny2017botnet} discusses how botnet resiliency can be enhanced using private blockchains. Pirozzi et al. \cite{Botchain} presented the use of blockchain as a command and control center for a new generation botnet. Kamenski et al. \cite{kamenski2019attacking} also show a proof of concept to build a Bitcoin-based botnet. Similarly, ChainChannels \cite{frkat2018chainchannels} utilizes Bitcoin to disseminate C\&C messages to the bots. These works are different from our architecture as they suffer from the issues of high latency and public announcement of commands. ZombieCoin \cite{ali2015zombiecoin} proposes to use Bitcoin's transaction spreading mechanism as the C\&C communication infrastructure. The authors later proposed ZombieCoin 2.0 \cite{ali2018zombiecoin} that employs subliminal channels to increase the anonymity of the botmaster. However, subliminal channels require a lot of resources to calculate required signatures which is computationally expensive and not practical to use on a large scale. A more recent work by Yin et al. \cite{yin2020coinbot} proposes CoinBot, a botnet that runs on cryptocurrency networks based on Bitcoin protocol such as Litecoin and Dash. Similar to other Bitcoin based botnets, CoinBot utilizes the \texttt{OP\_RETURN} field in the transactions. However, this approach is still costly and command propagation is slow.

There are also Unblockable Chains~\cite{UnblockableChains}, and Botract~\cite{Botract}, which are Ethereum \cite{wood2014ethereum} based botnet command and control infrastructures that suffer from anonymity issues since the commands are publicly recorded on the blockchain. Baden et al. \cite{whisper} proposed a botnet C\&C scheme utilizing Ethereum's Whisper messaging protocol. However, it is still possible to blacklist the topics used by the botmaster. Additionally, there is not a proof of concept implementation of the proposed approach yet, therefore it is unknown if the botnet can be successfully deployed or not.

The closest works to ours are the Franzoni et al. \cite{franzoni2020leveraging} and DUSTBot \cite{zhong2019dustbot}. Franzoni et al. propose to utilize the Bitcoin Testnet for controlling a botnet. Even though their C\&C communication is encrypted, non-standard Bitcoin transactions used for the communication exposes the botnet activity. Once the botnet is detected, the messages coming from the botmaster can be prevented from spreading, consequently stopping the botnet activity. DUSTBot also partially uses the Bitcoin Testnet in its design. Authors propose to utilize the Testnet as the upstream channel for sending the bot data back to the botmaster. However, botmaster's Bitcoin address is subject to blacklisting which blocks the communication of the botnet.

Finally, there are also botnets that utilize anonymity networks such as Tor \cite{brown2010resilient}. It is tempting for botmasters to use Tor and other anonymity networks due to their privacy and anonymity guarantees. However, there are numerous works in the literature showing ways to detect and stop botnets utilizing Tor. For example, Casenove et al. \cite{casenove2014botnet} showed that a botnet over Tor can be exposed due to the recognizable patterns in the C\&C communication. Sanatinia et al. \cite{sanatinia2015onionbots} proposed a Sybil mitigation technique to neutralize the bots in their proposed OnionBots which is a botnet utilizing Tor. Thus, as can be seen, anonymity networks are susceptible to other unique attacks associated with their inherent characteristics when used for botnet communication.

In contrast, our work is based on legitimate LN payments and does not require any additional computation to hide the commands. Also, these commands are not announced publicly. Moreover, LNBot offers a very unique advantage for its botmaster: C\&C servers do not have any direct relation with the botmaster thanks to LN's anonymous multi-hop structure. Even more, D-LNBot removes the costs that were present in LNBot and enables running a free botnet on LN. As a result, LNBot and D-LNBot do not carry any mentioned disadvantages through their two-layer hybrid architecture and provide superior scalability and anonymity compared to others. We would like to note that, a preliminary version of this paper that proposed LNBot was published in \cite{kurt2020lnbot}. This paper extends it with D-LNBot and iLNBot along with their design, implementation and analysis.

\section{Background}
\label{sec:background}

\subsection{Lightning Network}

The LN concept is introduced in \cite{poon2016bitcoin}. It is a payment protocol operating on top of Bitcoin. Through this concept, an overlay payment channel network (i.e., LN) is started among the customers and vendors in 2017. The aim in creating the LN was to decrease the load on the Bitcoin network, facilitating transactions with affordable fees and reduced transaction validation times, and thus increasing the scalability of Bitcoin \cite{mercan2021cryptocurrency, kurt2021lngate}. Despite the big fluctuations in the price of Bitcoin recently, the LN grew exponentially reaching 16,359 nodes and 73,580 channels in five years by the time of writing this paper\footnote{\url{https://1ml.com}}. In the following subsections, we briefly explain the components of LN. 

\subsection{Off-chain Concept}
\label{sec:offchainconcept}

\begin{figure*}[htb]
  \centering
  \includegraphics[width=0.75\linewidth]{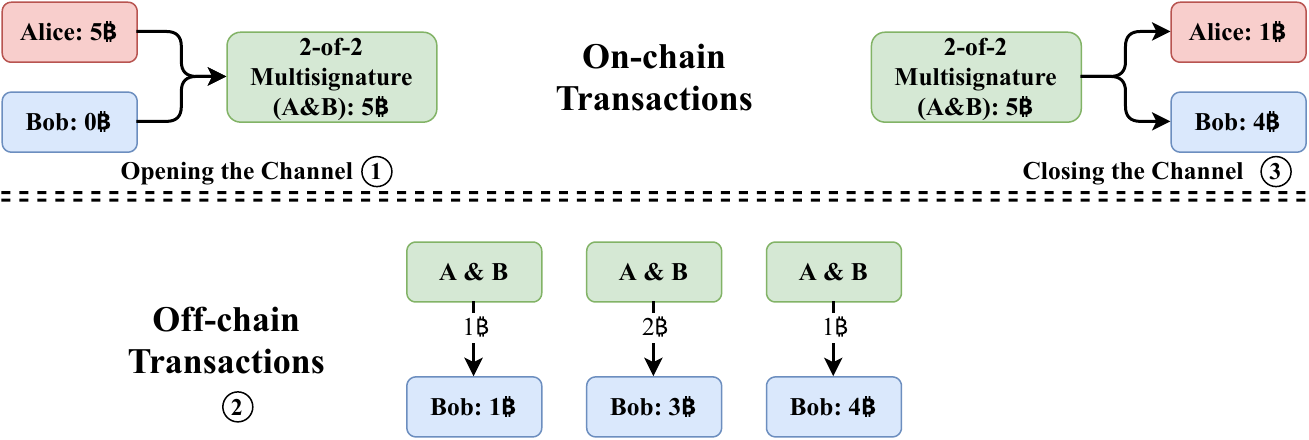}
     \vspace{-3mm}
  \caption{Off-chain mechanism of LN.}
  \label{fig:offchain}
  \vspace{-3mm}
\end{figure*}

The main idea behind LN is to utilize the \textit{off-chain} concept which enables near-instant Bitcoin transactions with negligible fees. This is accomplished through \textit{bidirectional payment channels} which are created among two parties to exchange funds without committing the transactions to Bitcoin blockchain. The channel is opened by making a single on-chain transaction called the \textit{funding transaction}. The funding transaction places the funds into a \textit{2-of-2 multisignature} address which also determines the capacity of the channel. Whenever the parties want to send a payment, they basically shift corresponding portion of their channel balance to the other side of the channel. So, after a transaction takes place, the total capacity in the channel does not change but the directional capacities do. Therefore, the peers can make as many transactions as they want in any amount as long as the amount they want to send does not exceed the directional capacity. The example shown in Fig. \ref{fig:offchain} illustrates the concept in more detail.

Per figure, \textcircled{{\scriptsize \textbf{1}}} Alice opens a channel to Bob by funding 5 Bitcoins into a 2-of-2 multi-signature address which is signed by both Alice and Bob.  \textcircled{{\scriptsize \textbf{2}}} Using this channel, Alice can send payments to Bob simply by transferring the ownership of her share in the multi-signature address until her funds in the address are exhausted. Note that these transactions are off-chain meaning they are not written to the Bitcoin blockchain which is a unique feature of LN that is exploited in our botnet. Alice performs 3 transactions at different times with amounts of 1, 2 and 1 Bitcoin respectively.  \textcircled{{\scriptsize \textbf{3}}} Eventually, when the channel is closed, the remaining 1 Bitcoin in the multi-signature address is returned to Alice while the total transferred 4 Bitcoins are settled at Bob. Channel closing is another on-chain transaction that reports the final balances of Alice and Bob in the multi-signature address to the blockchain.

\begin{figure}[htb]
  \centering
  \includegraphics[width=\linewidth]{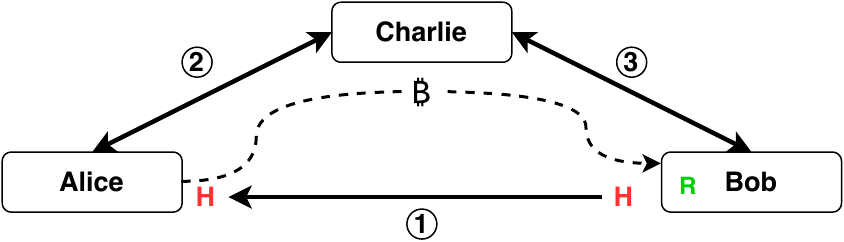}
  \vspace{-6mm}
  \caption{Illustration of a multi-hop payment. $R$ is the pre-image generated by Bob, $H$ is the hash of $R$. Alice creates an HTLC and includes $H$ in it. Through this mechanism, Alice can securely route her payment to Bob over Charlie.}
  \vspace{-3mm}
  \label{fig:multihop}
\end{figure}

\subsection{Multi-hop Payments}

In LN, a node can send payments to another node even if it does not have a direct payment channel to that node. This is achieved by utilizing already established payment channels between other nodes and such a payment is called multi-hop payment since the payment is forwarded through a number of intermediary nodes (i.e., hops) until reaching the destination. This process is trustless meaning the sender does not need to trust the intermediary nodes along the route. Fig. \ref{fig:multihop} depicts a multi-hop payment. Alice wants to send a payment to Bob but does not have a direct channel with him. Instead, she has a channel with Charlie which has a channel with Bob, thus Alice can initiate a payment to Bob via Charlie. For this to work, \textcircled{{\scriptsize \textbf{1}}} Bob first sends an \textit{invoice} to Alice which includes the hash ($H$) of a secret $R$ (known as \textit{pre-image}). \textcircled{{\scriptsize \textbf{2}}} Then, Alice prepares the payment through a contract called \textit{Hash Time Locked Contract} (HTLC) \cite{poon2016bitcoin} and includes $H$ inside the payment. HTLCs are used to ensure that the payments can be securely routed over a number of hops. Now, Alice sends this HTLC to Charlie and waits for Charlie to disclose the $R$. That is the condition for Alice to release the money to Charlie. \textcircled{{\scriptsize \textbf{3}}} In the same way, Charlie expects Bob to disclose $R$ so that he can send it to Alice the claim the money. Finally, when Bob releases the $R$ to Charlie, the payment will have successfully sent and HTLC will have fulfilled. Through this mechanism of LN, as long as there is a path from the source to the destination with enough liquidity, payments can be routed just like the Internet.

\subsection{Key Send Payments}
\label{sec:keysend}

\textit{Key send} in LN enables sending payments to a destination without needing to have an invoice first \cite{keysend}. It utilizes \textit{Sphinx}\cite{Sphinx} which is a compact and secure cryptographic packet format to anonymously route a message from a sender to a receiver. This is a very useful feature to have in LN because it introduces new use cases where payers can send spontaneous payments without contacting the payee first. In this mode, the sender generates the pre-image for the payment instead of the receiver and embeds the pre-image into the Sphinx packet within the outgoing HTLC. If an LN node accepts key send payments, then it only needs to advertise its public key to receive key send payments from other nodes. We utilize this feature to send payments from botmaster to C\&C servers in LNBot, iLNBot and D-LNBot.

\subsection{Source Routing, Onion Routed Payments and Private Channels}
\label{sec:onionrouting}
With the availability of multi-hop payments, a routing mechanism is needed to select the best route for sending a payment to its destination. LN utilizes \textit{source routing} which gives the sender full control over the route for the payment to follow within the network. Senders are able to specify: 1) the total number of hops on the path; 2) total cumulative fee they are willing to pay to send the payment; and 3) total time-lock period for the HTLC \cite{onionrouting2}. Moreover, all payments in LN are \textit{onion-routed payments} meaning that HTLCs are securely and privately routed within the network. Additionally, by encoding payment routes within a Sphinx packet, the following security and privacy features are achieved: the nodes on a routing path do not know: 1) the source and the destination of the payment; 2) their exact position within the route; and 3) the total number of hops in the route. Consequently, these features prevent any node from easily censoring or analyzing the payments.

In LN, it is optional to announce the opened channels publicly. The channels that are not publicly announced to the network are called \textit{private} channels. They are only known by the two peers of the channel. Other users cannot use the private channels to route payments over them. However, it is still possible to utilize these channels for routing payments with the help of \textit{routing hints}\footnote{\url{https://write.as/arshbot/everything-you-need-to-know-about-hop-hints-in-the-lightning-network}}. Routing hints are used to let the sender know how to reach an LN node behind a private channel.

\subsection{Attaching Messages to the Payments}
\label{sec:noiseplugin}
Recent developments in LN\footnote{\url{https://github.com/lightning/bolts/pull/619}} made it easier to attach arbitrary data to the payments. A protocol to enable this functionality was developed at 2019\footnote{\url{https://github.com/joostjager/whatsat}}. Later, the protocol was incorporated to the Core Lightning through the noise plugin \cite{noise}. We utilize the plugin for sending botmaster's commands to the C\&C servers in iLNBot and D-LNBot. Messages are included in the payload inside the onion packets that are routed over the hops until reaching the recipient. Onion size is 1300 bytes, thus theoretically, messages of size up to 1300 bytes can be sent in a single payment. This is much higher than the size allowed by the \texttt{OP\_RETURN} field in Bitcoin transactions which allows carrying only 80 bytes of data. The protocol carries the following information inside the onion payload: 1) key send pre-image, 2) message of variable size, 3) compressed signature and recovery id, and 4) timestamp. Intermediary nodes cannot the see payload inside the onion packet thus the messages can be anonymously sent to the recipients. Note that, different LN implementations might have a different version of this protocol or may not have it at all.

\section{Threat Model}
\label{sec:threatmodel}

The potential adversaries to the proposed botnets and to any botnet system in general include the governments, law enforcement, and security researchers who actively discover new botnets and develop methods to stop them. We assume that these parties can work with Internet Service Providers (ISPs) to track down IP addresses and physically locate devices. They possess access to advanced security and debugging software, including packet analyzers and tracing tools. Moreover, they are capable of reverse engineering the botnet malware, monitoring on-chain Bitcoin transactions, and deploying their own Bitcoin and LN nodes.

With these capabilities, attackers can perform various attacks, including compromising the bots and ultimately a C\&C server. They can also identify the hardcoded parameters within the botnet malware, conduct timing analysis on certain LN payments to detect the botmaster, send payments to C\&C servers to corrupt the commands, analyze on-chain payments generated by the botnet to expose the botmaster, analyze the network packets of certain LN payments, and attempt to blacklist specific nodes in the Bitcoin and LN networks by reaching out to Bitcoin and LN developers. Lastly, they might even attempt to reset Bitcoin's Testnet or completely shut down LN.

\section{LNBot Architecture}
\label{sec:LNBot}

In this section, we describe the overall architecture of LNBot with its elements.

\subsection{Overview}
\label{sec:LNBot_overview}

\begin{figure*}[h]
    \centering
    \includegraphics[width=0.8\linewidth]{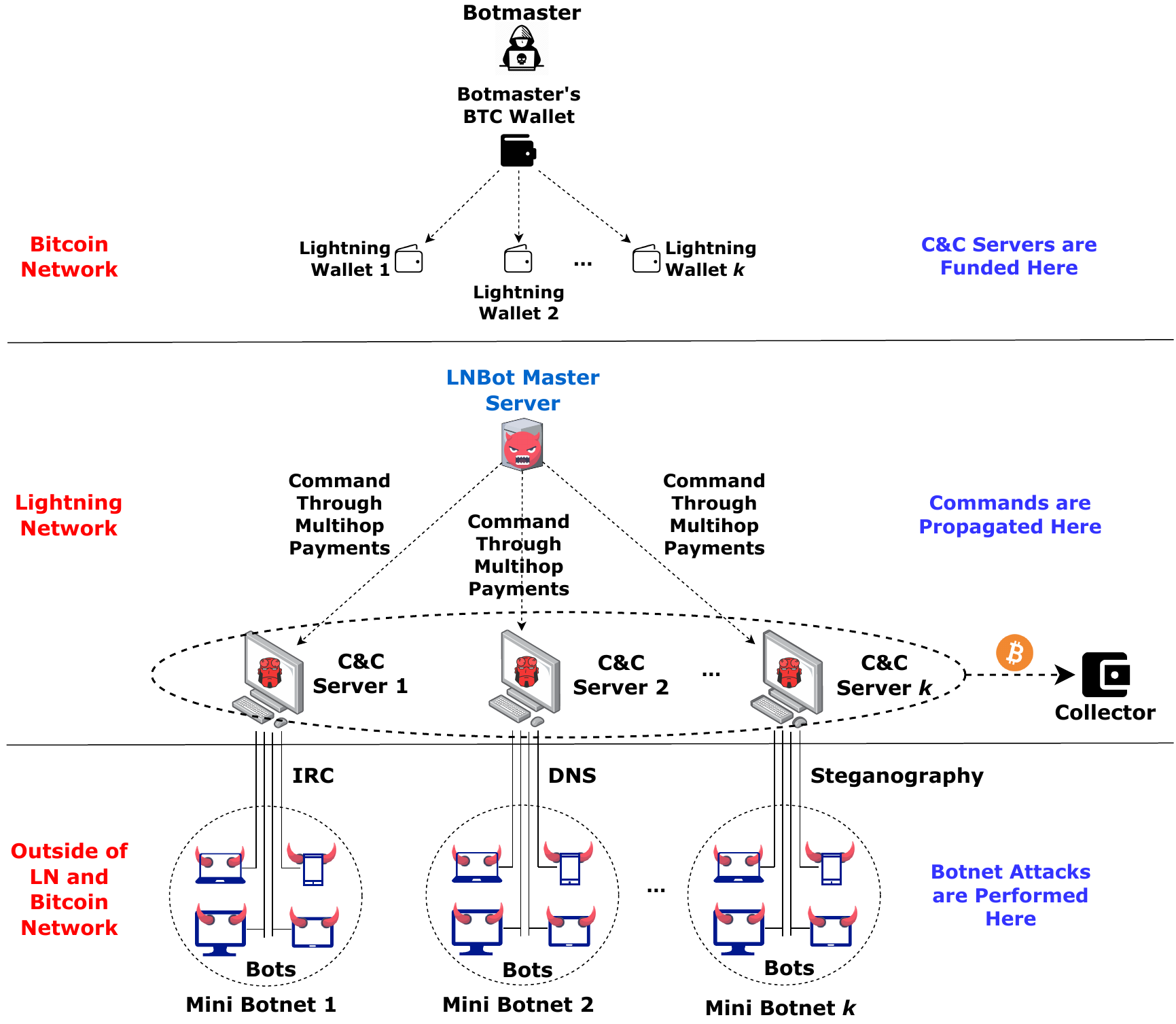}
    \vspace{-3mm}
    \caption{Overview of LNBot architecture.}
    \label{fig:lnbotoverview}
    \vspace{-4mm}
\end{figure*}

The overall architecture is shown in Fig. \ref{fig:lnbotoverview}. As shown, the LN is used to maintain the C\&C servers and their communication with the botmaster. Each C\&C server runs a separate mini-botnet. Receiving the commands from the botmaster, the C\&C servers relay the commands to the bots in their possession to launch attacks. The details of setting up the C\&C servers are explained next.

\subsection{Setting up the C\&C Servers}
The botmaster sets up the necessary number of C\&C servers s/he would like to deploy. Depending on the objectives, the number of these servers and the number of bots they will control can be adjusted without any scalability concern. In Section \ref{sec:POC}, we explain how we set up real C\&C servers running on LN on the real Bitcoin network.

Each C\&C server is deployed as a \textit{private} LN node which means that they do not advertise their channels within the LN. This helps C\&C servers to stay anonymous in the network. Each C\&C server opens private channels to at least $k$ different random public LN nodes. The number $k$ may be tuned depending on the size and topology of LN when LNBot will have deployed in the future. To open the channels, C\&C servers need some Bitcoin in their Bitcoin wallets. This Bitcoin is provided to the C\&C servers by the botmaster before deploying them. Since C\&C servers' channels are unannounced (i.e., private) in the network, botmaster uses \textit{routing hints} to be able to route his/her payments to the C\&C servers. In this case, since the C\&C servers are set up by the botmaster, s/he knows the necessary information to create the routing hints.

\subsection{Formation of Mini-botnets}
After the C\&C servers are set up, we need bots to establish connections to the C\&C servers. An infected machine (bot) connects to one of the C\&C servers available. The details of bot recruitment and any malware implementation issues are beyond the objectives of this paper. It is up to the botmaster to decide which type of infrastructure the C\&C servers will use to control the bots in their possession. This flexibility is enabled by our proposed two-layer hybrid architecture of the LNBot. The reason for giving this flexibility is to enable scalability of LNBot through any type of mini-botnets without bothering for the compromise of any C\&C servers. As it will be shown in Section \ref{sec:countermeasures}, even if some of the C\&C servers are compromised, this neither reveals the other C\&C servers nor the botmaster.

\subsection{Forming LNBot}
Now that the C\&C servers are set up and mini-botnets are formed, the next step is to form the infrastructure to control these C\&C servers covertly with minimal chances of getting detected. This is where LN comes into play. Botmaster has the LN public keys of all the C\&C servers. Using an LN node called \textit{LNBot master server}, the botmaster initiates the commands to all the C\&C servers through LN payments. Similar to the C\&C servers, LNBot master server is also a private LN node and the botmaster has flexibility on the setup of this node and may change it regularly. Without using any other custom infrastructure, the botmaster is able to control the C\&C servers through LN, consequently controlling all the bots in the botnet.

\subsection{Command Propagation in LNBot}
\label{sec:LNBot_com_prop}

Once the LNBot is formed, the next step is to ensure communication from the botmaster to the C\&C servers. We utilize a \textit{one-to-many} architecture where the botmaster sends the commands to each C\&C server separately. Commands can be sent in two different ways: 1) Encoding the characters in the command to individual LN payments; 2) Attaching the whole command to a single LN payment. The first method is utilized in LNBot and the second one is utilized in \textit{iLNBot}. These two methods are explained separately at the sections below.

\subsubsection{Encoding/Decoding Schemes}
\label{sec:LNBot_encodings}
An important feature of LNBot is its ability to encode botmaster's commands into a series of LN payments that can be decoded by the C\&C servers. We used ASCII encoding and Huffman coding \cite{huffman2006method} for the purpose of determining the most efficient way of sending commands to the C\&C servers in terms of Bitcoin cost and time spent. American Standard Code for Information Interchange (ASCII) is a character encoding standard that represents English characters as numbers, assigned from 0 to 127. Huffman coding on the other hand is one of the optimal options when the data needs to be losslessly compressed. We used Quaternary numeral system to generate the codebook as will be shown in Section \ref{sec:evaluation}.

\SetKw{KwIn}{in}
\begin{algorithm}[h]
\label{algo:algorithm1}
\SetAlgoLined
define $k$\;
initialize $command$\;
int $counter$ = 0\tcc*{retry count}
bool $isOnline$ = checkIfC\&CServerIsOnline()\;
\uIf {$isOnline$}{
    bool $result$ = send($5$ $satoshi$)\;
    \uIf {$result$==success}{
        $counter$ = 0\;
        \For{$character$ \KwIn $command$}{
            bool $result$ = send($character$)\;
            \lIf {$result$==success}{continue}
            \uElseIf {$result$==fail and $counter<k$} 
                {retry sending $character$\;
                $counter$++;}
            \lElse {reschedule($command$, date, time)}
        }
        $counter$ = 0\;
        bool $result$ = send($6$ $satoshi$)\;
        \uIf {$result$==success}
            {\textbf{Command has been successfully sent!};}
        \uElseIf{$result$==fail and $counter<k$}
            {retry sending $6$ $satoshi$\;
            $counter$++;}
        \lElse {reschedule($command$, date, time)}
    }
    \uElseIf{$result$==fail and $counter<k$}
        {retry sending $5$ $satoshi$\;
        $counter$++;}
    \lElse{reschedule($command$, date, time)}
}
\Else{reschedule($command$, date, time);}
\caption{Send Command}
\end{algorithm}

For both methods, botmaster uses the Algorithm \ref{algo:algorithm1} to send the commands. S/he first checks if the respective C\&C server is online or not (LN nodes have to be online in order to send and receive payments) before sending any payment. If the C\&C server is not online, command sending is scheduled for a later time. If online, the botmaster sends 5 satoshi as the special starting payment of a command, then the actual characters of the command one by one. Lastly, the botmaster sends 6 satoshi as the special ending payment to finish sending the command. Note that the selection of 5 satoshi and 6 satoshi in this algorithm depends on the chosen encoding and could be changed based on the needs. If any of these separate payments fail, it is retried. If any of the payments fail for more than $k$ times in a row, command sending to the respective C\&C server is canceled and scheduled for a later time.

\subsubsection{Noise Plugin Method}
\label{sec:lnbot_w_noise}
When LNBot was introduced, attaching custom data to LN payments was a work in progress by the LN developers. As explained in Section \ref{sec:noiseplugin}, noise plugin \cite{noise} gives one the opportunity to embed messages in LN payments. If we utilize this plugin in LNBot for sending botmaster's commands, the command sending will be much faster and cheaper. Specifically, the delay of sending a command to a C\&C server will be reduced to a delay of sending a single payment. In the same way, the cost of sending a command will now be the cost of sending a single payment. Thus, we propose \textit{iLNBot} which utilizes the noise plugin as its command sending mechanism instead of the encoding method presented earlier.

\subsection{Reimbursing the Botmaster}

Another important feature of LNBot is the ability of its botmaster to get the funds back from the C\&C servers. Depending on botmaster's command propagation activity, C\&C servers' channels will fill up with the funds received from the botmaster. Therefore, in our design, C\&C servers are programmed to send the funds in their channels to an LN node called \textit{collector} when their channels fill up completely. Collector is a private LN node set up by the botmaster. Its LN public key is stored in the C\&C servers and thus they can send the funds to collector through LN. In addition to collector's LN public key, C\&C servers are also supplied with routing hints to be able to successfully route their payments to the collector. In this way, the funds accumulate at the collector. The botmaster can collect these funds by closing collector's respective LN channels and sending the settled Bitcoins to his/her own Bitcoin wallet through on-chain transfers.

\section{D-LNBot Architecture}
\label{sec:D-LNBot}

The LNBot we presented in the previous section relies on a centralized model where the botmaster communicates with each C\&C server separately. Additionally, in LNBot, C\&C servers are manually set up by the botmaster which incurs several operating costs on the botmaster that grows with each installed C\&C server. Specifically, botmaster needs to create machines to run the C\&C servers and the LN nodes on them as well as fund each of these LN nodes which costs time and money. On top of that, LNBot's centralized design might tamper the anonymity of the botmaster and the C\&C servers.   
Therefore, in this section, we present a distributed version of LNBot, namely D-LNBot, where the payments sent by the botmaster are reduced to a single transaction (i.e., to a specific C\&C server) regardless of the size of the botnet. On top of the distributed design, we propose to utilize the Bitcoin Testnet instead of the Mainnet which removes the associated costs of running the botnet.

\subsection{Overview}
\label{sec:Overview}

\begin{figure}[h]
    \centering
    \includegraphics[width=\linewidth]{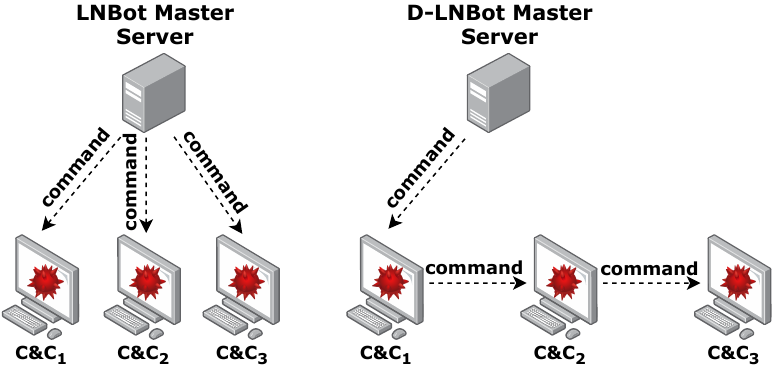}
    \vspace{-6mm}
    \caption{Comparison of the command propagation between LNBot and D-LNBot.}
    \label{fig:d-lnbot-overview}
    \vspace{-4mm}
\end{figure}

In this section, we describe the overall architecture of D-LNBot with its elements. LN is used for the communication between the botmaster and the C\&C servers also among the C\&C servers themselves. Different than LNBot, in D-LNBot, botmaster does not send the commands to each C\&C server individually. Rather, the commands are just sent to one of the C\&C servers which relays them to all other C\&Cs distributively as will be explained in the subsequent sections. We illustrated this difference in command sending between the LNBot and D-LNBot in Fig. \ref{fig:d-lnbot-overview}. We propose to utilize the noise plugin for sending the commands. Similar to the LNBot, each C\&C server controls a mini-botnet which can utilize any existing known C\&C infrastructure in the literature.

\subsection{Creation of the C\&C Servers}

The way that the C\&C servers are created differs from the LNBot. Botmaster infects existing machines on the Internet to turn them into C\&C servers. Thus, s/he deploys a malware into the wild for this purpose. Deployment details of this malware is out of the scope of this paper. However for later reference, we will call it the \textit{malware\_1}. Once a machine is infected with \textit{malware\_1}, there are two possibilities: 1) With probability \textit{p}, the machine becomes a C\&C server or 2) with probability \textit{1-p}, nothing happens and the malware deletes itself. Ultimately, the probability \textit{p} should be less than 0.5 to ensure that C\&C servers are generated occasionally. The reason for not turning every infected machine into a C\&C server is to limit the number of C\&C servers that will be ever created. If botmaster lets every infected machine to turn into a C\&C server, that will increase the number of public LN nodes in LN on Bitcoin's Testnet dramatically. As of May 2023, LN on Bitcoin's Testnet have around 2,400 public nodes\footnote{\url{https://1ml.com/testnet/}}. Considering how fast a malware can infect new machines, we propose that it is best to limit the number of C\&C servers that can be generated with the malware. There might be several ways to do this and we leave its design to the botmaster. After this process, if the machine turns into a C\&C server, it will download an LN software first. We propose to go for an LN light client instead of a full LN node since downloading a full LN node may alert the user of the infected machine as the download size is around 480 GB at the time of writing this paper. Light LN clients such as Neutrino \cite{neutrino} occupy a very small space (around 1 MB) on the file system and have higher chances of going unnoticed by the users. After the LN node is created and initiated, it needs to be funded with some Testnet Bitcoins. For this, we suggest to use a pre-funded Testnet Bitcoin wallet where the C\&C server can fetch the Bitcoins from. Generally, Testnet Bitcoins are obtained from Testnet faucets\footnote{e.g., \url{https://testnet-faucet.mempool.co/}}. The procedure of pre-funding a wallet can be tailored by the botmaster to leave as little on-chain trace as possible. After the C\&C server obtains some Testnet Bitcoins, the next step is the recruitment of the bots to work under the C\&C server (i.e., formation of the mini-botnet).

\subsection{Formation of the Mini-botnets}

Similar to the LNBot, the botmaster uses a specific malware for this purpose. Let us call it \textit{malware\_2}. The machines that are infected with \textit{malware\_2} become bots and connect to the available C\&C servers. When enough number of machines are infected with \textit{malware\_2}, they form a mini-botnet which are controlled by their respective C\&C server. Here, the command and control infrastructure to control the bots can be different for each mini-botnet. There are many available C\&C types such as DNS, social media, IRC, blockchain etc. \cite{vormayr2017botnet}. Botmaster should include as many of these C\&C channel options as possible into the \textit{malware\_2} so that the C\&C servers can randomly utilize one of them. These randomization is required since in D-LNBot, the C\&C servers are not set up by the botmaster. Otherwise, botmaster would have to create variations of \textit{malware\_2} which utilize different C\&C channels. However, this creates additional management overhead for the botmaster therefore we opt to create only one malware for this purpose that includes different C\&C channel options where C\&C servers can randomly choose from. This flexibility is a design choice which enables creating a \textit{hybrid botnet}. So, ultimately botmaster deploys two malware into the wild, namely \textit{malware\_1} and \textit{malware\_2}, to create the C\&C servers then to create bots to work under them. Next step is to create the connections between the C\&C servers to form the D-LNBot.

\subsection{Forming D-LNBot using Innocent Nodes}
\label{sec:dlnbot_forming}

An important challenge in D-LNBot is to piece together the D-LNBot from individual C\&C servers without the botmaster intervening in the process. We propose forming the communication among the C\&C servers by exploiting the connections that can be created with existing nodes in LN. We will call them \textit{innocent nodes} and basically these are LN nodes that are publicly reachable by other nodes. Current LN implementation lets users explore the network topology. For example, typing \texttt{lncli describegraph} (\texttt{lncli} is \textit{lnd's} \cite{lnd} command line interface) returns the network topology information in JSON format that can be analyzed to get information about the nodes reachable from our node. The information that can be collected include their public keys, IP addresses, channels, and channels' capacities.

\begin{figure*}[h]
    \centering
    \includegraphics[width=0.9\linewidth]{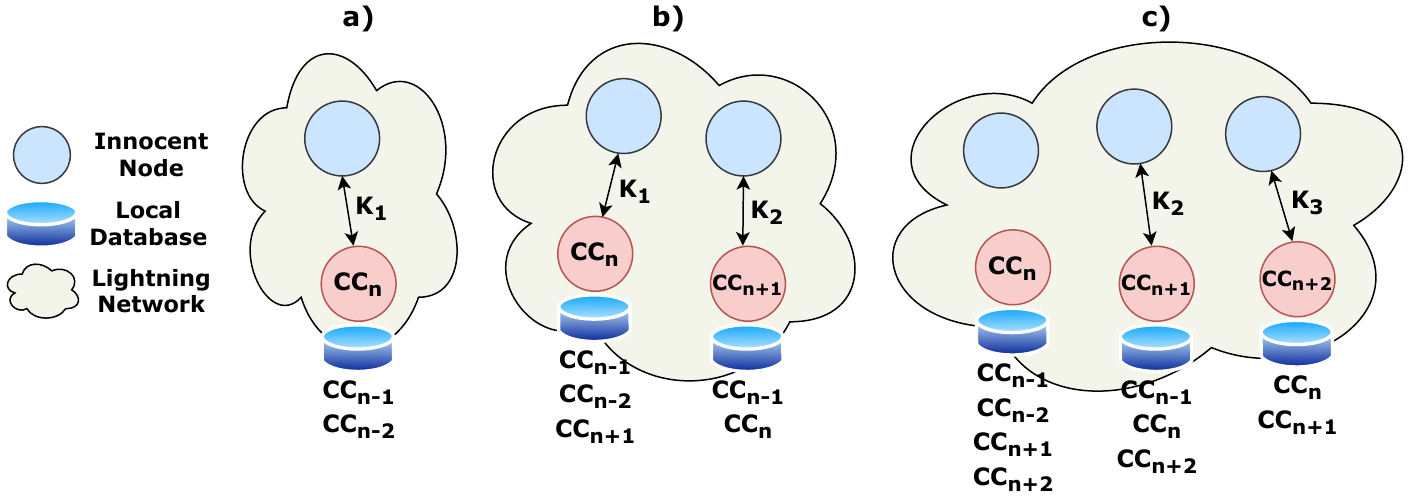}
    \vspace{-4mm}
    \caption{An illustration of how D-LNBot is formed when number of active C\&C servers is 2 (i.e., $m=2$). a) C\&C$_n$ joins the network and opens a channel to an innocent node with a capacity $K_1$. b) C\&C$_{n+1}$ joins the network and opens a channel to an innocent node with a capacity $K_2$. c) C\&C$_{n+2}$ joins the network and opens a channel to an innocent node with a capacity $K_3$ which results in C\&C$_n$ closing its channel with the innocent node. The process of C\&C servers registering each other as neighboring servers into their local databases are also illustrated.}
    \label{fig:D-LNBot-formation}
    \vspace{-4mm}
\end{figure*}

The steps of forming the D-LNBot is shown in Algorithm \ref{algo:algorithm2} as well as illustrated in Fig. \ref{fig:D-LNBot-formation}. First step is to identify some innocent nodes. When a new C\&C server, say C\&C$_n$, is created, it will establish channels with some of these innocent nodes following a specific policy. To do that, first, C\&C$_n$ queries the LN and finds the $h$ most connected nodes in the network. Next, among these $h$ nodes, C\&C$_n$ randomly selects one of them and opens a channel to it with a capacity $K_1$ (line 7 in Algorithm \ref{algo:algorithm2}). As for choosing the value of $K_1$, it should satisfy the following condition: $f(K_1)=\xi$, where $f()$ is a general function and $\xi$ is an unique value (line 3-6). Then, C\&C$_n$ scans the LN to discover the existing C\&C servers (line 8). This is easy to do by querying the LN and checking the capacity of each channel to see if it is satisfying the rule given earlier. As soon as C\&C$_n$ finds a channel satisfying the rule, it will register the node that opened this channel as a C\&C server to its local database (line 16). Here, even though discovering the existing C\&C servers is the goal, it is not desirable to reveal the presence of all existing C\&C servers at once. Therefore, older C\&C servers should hide their existence and the newly joined C\&C servers should be able to discover only a small number of C\&C servers at any given time. This is possible by older C\&C servers closing their open channels with the innocent nodes after getting discovered by $m$ new C\&C servers (line 31). For example, if $m$ is set to 2, C\&C$_1$ will close its open channel with the innocent node after getting discovered by the C\&C$_3$. C\&C$_2$ will do the same after C\&C$_4$ joins the network. In this way, there will always be $m$ C\&C servers only that are discoverable by a newcomer at any time. This number can be adjusted based on the needs and we will refer to it as the \textit{number of active C\&C servers}. But, how can the existing C\&C servers know that a new C\&C server joined the network? This is possible by existing C\&C servers monitoring each new channel creation on LN and checking if the channel capacity satisfies the given rule. New channels are already announced to the network automatically so one would only need to implement a helper function to detect the channels that were created by the newly joined C\&C servers (line 10). Following this mechanism, each C\&C server will be aware of $m$ older C\&Cs and $m$ newly joined C\&Cs (except for the first $m$ C\&C servers).

Next, assume that a new C\&C server, C\&C$_{n+1}$, joins the network. Similarly, it will establish a channel with an innocent node with a capacity $K_2$ where $f(K_2)=\xi$ and query the LN to find the older C\&Cs. It will find out that, C\&C$_n$ has a channel with an innocent node that satisfies the condition $f(K_1)=\xi$. Thus, C\&C$_{n+1}$ will add C\&C$_n$ as a C\&C server to its local database (line 16). In the mean time, C\&C$_n$ will observe that C\&C$_{n+1}$ has joined LN since it opened a channel satisfying the rule $f(K_2)=\xi$ and will add C\&C$_{n+1}$ to its local database (line 28). This procedure goes on as new C\&C servers join the network. Here, the function $f()$ and value $\xi$ can be hard-coded in the \textit{malware\_1}. An example function for $f()$ could be $sin()$. One possible issue that might arise is that when new C\&C servers join the network, the $h$ most connected nodes in LN might change. This might cause the newly joined C\&C servers to not be able to discover the existing C\&C servers and vice versa. To remedy this issue, the C\&C servers that could not discover at least two other C\&C servers repeat the process of opening a channel to an innocent node (line 20). This will make sure that all C\&C servers are aware of at least two other C\&C servers. Essentially, this process forms an \textit{overlay network} on top of LN to control a botnet.

\begin{algorithm}[h]
\caption{Form D-LNBot}
\label{algo:algorithm2}
\SetAlgoLined
\textbf{define} $h$, $m$, $f()$, $\xi$\;
\textbf{global} $K$ = []\tcc*{an empty list}
\For{$i\leftarrow 1$ \KwTo $n$}{
    \tcc{calculate $n$ valid channel capacities for the policy based on $f()$ and $\xi$}
    solve $f(K_i)=\xi$ for $K_i$\;
    $K$.append($K_i$)\;
} 
\textbf{global} $node$ = \texttt{open\_channel\_to\_innocent()}\;
\texttt{check\_for\_existing\_C\&Cs()}\;
\textbf{global} $counter$ = 0\tcc*{newcomer C\&C count}
\texttt{monitor\_new\_channels()}\;

\SetKwFunction{FMain}{check\_for\_existing\_C\&Cs}
 \SetKwProg{Fn}{Function}{:}{end}
   \Fn{\FMain{void}}{
        $LN\_topology$ = query\_LN()\;
        $count$ = 0\tcc*{existing C\&C count}
        \For{$channel$ \KwIn $LN\_topology$}{
            \If{$f(channel.capacity) == \xi$}{
                \textbf{register} $channel$.getNode()\;
                $count$++\;
            }
        }
        \lIf{$count < 2$}{\texttt{open\_channel\_to\_innocent()}}
  }

\SetKwFunction{FMain}{monitor\_new\_channels}
 \SetKwProg{Fn}{Async Function}{:}{end}
   \Fn{\FMain{void}}{
        \tcc{will catch each new $channel$}
        \KwRet \texttt{is\_C\&C}($channel$)\; 
  }
  
\SetKwFunction{FMain}{is\_C\&C}
 \SetKwProg{Fn}{Function}{:}{end}
   \Fn{\FMain{$channel$}}{
        \If{$f(channel.capacity) == \xi$}{
            \uIf{$counter < m$}{
                \textbf{register} $channel$.getNode()\;
                $counter$++\;
            }
            \Else{
                close\_channel($node$)\;
            }    
        }
  } 

\SetKwFunction{FMain}{open\_channel\_to\_innocent}
 \SetKwProg{Fn}{Function}{:}{end}
   \Fn{\FMain{void}}{
        $innocent\_nodes$ = []\tcc*{an empty list}
        $LN\_topology$ = query\_LN()\;
        $innocent\_nodes$ = get\_h\_most\_connected($LN\_topology$, $h$)\;
        $node$ = randomly\_select\_one($innocent\_nodes$)\;
        $K_1$ = randomly\_select\_one($K$)\;
        open\_channel($node$, $K_1$)\;
        \KwRet $node$\;
    } 
\end{algorithm}

\subsection{Command Propagation in D-LNBot}
\label{sec:dlnbotcommandpropagation}

\begin{figure}[h]
    \centering
    \includegraphics[width=0.9\linewidth]{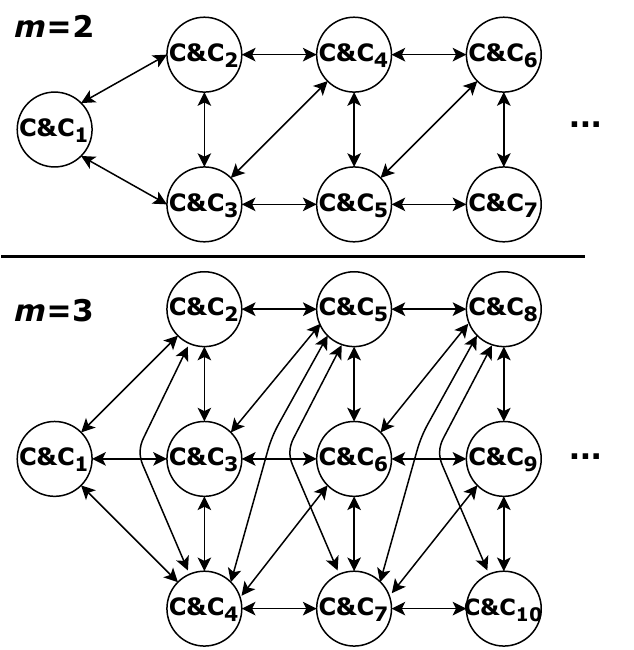}
    \vspace{-3mm}
    \caption{An illustration of the logical topology and command propagation of two sample D-LNBots when the number of active C\&C servers is two and three respectively (i.e., $m=2$ and $m=3$). In this example, botmaster initiates the command sending from C\&C$_1$.}
    \label{fig:commandProp}
    \vspace{-3mm}
\end{figure}

Unlike LNBot, D-LNBot utilizes a \textit{logical topology} created during the setup for command propagation. To initiate the command propagation, the botmaster sends the command from a node called \textit{D-LNBot master server} to a particular C\&C server which sends it to the C\&C servers it recorded in its local database. We will call these recorded C\&Cs, \textit{neighboring C\&C servers}. Every C\&C server does the same so the command is propagated among the C\&C servers in a P2P fashion. We illustrate two logical D-LNBot topologies in Fig. \ref{fig:commandProp} when the number of active C\&C servers is two and three. The bidirectional links show the command propagation between the respective C\&C servers. Note that, the C\&C servers are not connected to each other with direct channels.

For the $m = 2$ case in Fig. \ref{fig:commandProp}, the propagation starts when C\&C$_1$ receives a command from the botmaster. C\&C$_1$ immediately relays the command to its neighbors C\&C$_2$ and C\&C$_3$ by sending them a single payment which includes the command. Then, when C\&C$_2$ receives the command, it also immediately forwards the command to its neighbors C\&C$_1$, C\&C$_3$, and C\&C$_4$. C\&C$_3$ does the same when it receives the command from C\&C$_1$ or C\&C$_2$. Here, each C\&C server sends and receives the command multiple times but this do not result in multiple executions of the same command. Because, each C\&C server knows its neighbors and upon receiving a command from one of its neighbors, the C\&C server can ignore the command if it was received from another neighbor before. This requires being able to authenticate the sender of the commands which the C\&C servers can do. As mentioned in Section \ref{sec:noiseplugin}, the noise plugin messages include a \textit{signature}. Using the signature, the LN public key of the sender can be recovered. This means that, a C\&C server can compare this public key with the ones it has in its local database to make sure the command is coming from a legitimate C\&C server. Additionally, the redundancy on sending and receiving the commands helps C\&C servers keep their channels balanced. For the command propagation of $m = 3$ case, please refer to Fig. \ref{fig:commandProp}.

The redundancy on command propagation also helps detecting any false positives during the D-LNBot formation phase. Let us imagine a scenario where an honest LN node opens a channel with a capacity satisfying the policy used by the C\&C servers. This will be automatically detected by the active C\&C servers (there are only \textit{m} of them) who were actively querying the LN to detect the newly joined C\&C servers. The active C\&C servers will register this honest LN node to their local database. Now, let us imagine that the botmaster issues a command which starts getting propagated among the C\&C servers which will eventually reach the C\&Cs that registered the honest node as a neighbor C\&C. At that point, the honest node will receive a number of noise messages from these C\&Cs but will not propagate anything back. In fact, it might not even receive the messages successfully because of the custom TLV (type-length-value) used by the noise messages. Regardless, when the C\&Cs do not get the command back from the honest node, they can understand it was not actually a C\&C server thus, mark it as a false positive and delete it from their local database.

One advantage of D-LNBot is that, the botmaster can initiate the command propagation from any one of the C\&C servers because the initial C\&C server will replicate the command to its neighboring C\&C servers. In other words, the botmaster does not need to initiate the command propagation from C\&C$_1$ necessarily, and can use the C\&C$_5$ instead for example. Botmaster is able to do this because s/he is aware of all the C\&C servers as s/he is constantly watching the network and knows the policy the C\&C servers use to open channels to the innocent nodes. Being able to freely choose the first C\&C server to initialize the command propagation also provides better anonymity for the botmaster.

\section{Proof-of-Concept Implementation}
\label{sec:POC}

In this section, we demonstrate that implementations of the proposed LNBot and D-LNBot are feasible by presenting a proof-of-concept for each. Full LN nodes interact with the Bitcoin network in order to run the layer-2 protocols. There are two Bitcoin networks: \textit{Bitcoin Mainnet} and \textit{Bitcoin Testnet}. As the names suggest, Bitcoin \textit{Mainnet} hosts Bitcoin transfers with a real monetary value. On the contrary, in Bitcoin \textit{Testnet}, Bitcoins do not have a monetary value. They are only used for testing and development purposes. Nonetheless, they both provide the same infrastructure and LNBot and D-LNBot can run on both networks. The only difference between the Testnet and Mainnet networks for LN is the number of nodes and the channels they have. LN on Bitcoin Testnet have about 60\% less nodes, and 80\% less channels compared to LN on Bitcoin Mainnet.

\vspace{1mm}
\textbf{LNBot:} We used \textit{lnd} (version 0.9.0-beta) from Lightning Labs \cite{lnd} for the full LN nodes. We used \textit{Bitcoin Testnet} for our proof-of-concept development. We created 100 C\&C servers and assessed certain performance characteristics for command propagation. We created a GitHub page explaining the steps to set up the C\&C servers\footnote{\url{https://github.com/LightningNetworkBot/LNBot}}. The steps include installation of \textit{lnd} \& \textit{bitcoind}, configuring \textit{lnd} and \textit{bitcoind}, and extra configurations to hide the servers in the network by utilizing private channels. Nevertheless, to confirm that the channel opening costs and routing fees are exactly the same in both Bitcoin \textit{Mainnet} and \textit{Testnet}, we also created 2 nodes on Bitcoin \textit{Mainnet}. We funded one of the nodes with 0.01 Bitcoin, created channels and sent payments to the other node. We observed that the costs and fees are exactly matching to that of Bitcoin \textit{Testnet}.

\textit{lnd} has a feature called \textit{autopilot} which opens channels in an automated manner based on certain initial parameters set in advance\footnote{\url{https://github.com/lightningnetwork/lnd/blob/master/sample-lnd.conf}}. Our C\&C servers on Bitcoin \textit{Testnet} employ this functionality of \textit{lnd} to open channels on LN. Using \textit{autopilot}, we opened 3 channels per server. Note that this number of channels is picked based on our experimentation on Bitcoin \textit{Testnet} on the success of payments. We wanted to prevent any failures in payments by tuning this parameter.  As mentioned, these 3 channels are all private, created with \texttt{--private} argument, which do not broadcast themselves to the network. A private channel in LN is only known by the two peers of the channel.

\textit{lnd} has an API for communicating with a local \textit{lnd} instance through gRPC\footnote{\url{https://lightning.engineering/api-docs/api/lnd/}}. Using the API, we wrote a client that communicated with \textit{lnd} in Python. Particularly, we wrote two Python scripts, one running on the C\&C servers and the other on the botmaster machine. We typed the command we wanted to send to the C\&C servers in a terminal in the botmaster machine. The command was processed by the Python code and sent to the C\&C servers as a series of payments.

\vspace{1mm}
\textbf{D-LNBot:} Different than LNBot, we used \textit{Core Lightning} (version 0.10.0) for the full LN nodes in D-LNBot. It is one of the implementations of LN developed by Blockstream \cite{clightning}. The details of setting up a Core Lightning node that is configured to run on the C\&C servers are given at our GitHub page\footnote{\url{https://github.com/startimeahmet/D-LNBot}}. Specifically, we explain the steps for installing \textit{Core Lightning} and \textit{bitcoind}, configuring \textit{Core Lightning} and \textit{bitcoind}, and installing the necessary \textit{Core Lightning plugins}. Similar to the LNBot, we used Bitcoin's Testnet for our proof-of-concept development. We created 3 C\&C servers at various locations in US. The channels for the C\&C servers were opened manually instead of using the \textit{autopilot} feature of LN. The number of opened channels were 3 for all C\&C servers. To send the botmaster's commands, we utilized the \textit{noise} plugin in \textit{Core Lightning}. Basically, it enables one to attach messages to the key send payments. Since the commands are sent using a single payment, there was no need for writing a script to serialize the payments into commands like in LNBot.

In addition to setting up C\&C servers on Bitcoin networks, we performed several simulations in a specially crafted time-driven simulation environment using Python. In our setup, we simulated the D-LNBot formation phase and the command propagation in D-LNBot. The simulation environment can be found on our GitHub repository as well. Specifically, in the simulations, C\&C servers joined the network at random times and attempted to discover each other based on a certain policy as explained in Section \ref{sec:dlnbot_forming}. For the command propagation, using the generated network, the C\&C servers sent the commands to their neighbors in a P2P fashion as explained in Section \ref{sec:dlnbotcommandpropagation}. In these simulations, we used an up-to-date LN topology which is made publicly available \cite{lngossip}. The dataset is a collection of LN's gossip messages that is collected by a number of LN nodes running on Bitcoin Mainnet.

\section{Evaluation and Analysis}
\label{sec:evaluation}

In this section, we present a detailed cost and delay analysis of LNBot, iLNBot, and D-LNBot. While analyzing the different implementations, it is important to note that the selection of Bitcoin Mainnet and Bitcoin Testnet dramatically affects the evaluation results. While the infrastructure they offer are exactly the same, the cost of running the botnet is different. Testnet Bitcoins do not have a monetary value and thus they can be obtained for free from the faucets. On the other hand, Mainnet Bitcoins do have a real monetary value and have to be purchased.

\subsection{LNBot Evaluation}
While it is mostly a design choice, we used Bitcoin Mainnet to evaluate the LNBot because of its bigger size. Also, as will be explained in Section \ref{sec:countermeasures}, historically Mainnet had more uptime than Testnet.

\subsubsection{Cost Analysis of LNBot Formation}
\label{sec:LNBot_formation}

We first analyze the monetary cost of forming LNBot. As noted earlier, we opened 3 channels per server. The capacity of each channel is 20,000 satoshi which is the minimum allowable channel capacity in \textit{lnd}. Therefore, a server needs 60,000 satoshi for opening these channels. While opening the channels, there is a small fee paid to Bitcoin miners since channel creations in LN are on-chain transactions. We showed that, opening a channel in LN can cost as low as 154 satoshi on both Bitcoin Testnet\footnote{Check LNB6's channel opening transaction for instance: \url{https://blockstream.info/testnet/tx/fc46c99233389d24c4fd9517cd503f08265c517a6f0570d806e7cc98b7f7963b}} and the Mainnet\footnote{In a similar way, check one of our mainnet nodes' channel opening transaction: \url{https://blockstream.info/tx/1d81b6022ff1472939c4db730ca01b82d43b616e757d799aea17ee0db6427520}}.

So the total cost of opening 3 channels for a C\&C server is 60,462 satoshi. While 462 satoshi is consumed as fees, the remaining 60,000 satoshi on the channels is not spent, rather it is just locked in the channels. The botmaster will get this 60,000 satoshi back after closing the channels. Therefore, funds locked in the channels are non-recurring investment cost for the formation of LNBot. Only real associated cost of forming LNBot is the channel opening fees.

Table \ref{tab:channelopening} shows how the costs change when the number of C\&C servers is increased. The increase in the cost is linear and for 100 C\&C servers, the on-chain fees are only 46,200 satoshi.

\begin{table}[htb]
\vspace{-3mm}
  \begin{center}
    \caption{Channel Opening Fees for Different Number of C\&C Servers}
    \vspace{-3mm}
    \label{tab:channelopening}
    \resizebox{0.87\linewidth}{!}{
    \begin{tabular}{|c|c|c|c|}
    \hline
      \textbf{Number of C\&C Servers} & \textbf{Channel Opening Fees}  \\
      \hline
      10 & 4,620 satoshi \\
      \hline
      25 & 11,550 satoshi \\
      \hline
      50  & 23,100 satoshi \\
      \hline
      100 & 46,200 satoshi \\
      \hline
    \end{tabular}
  }
  \vspace{-5mm}
  \end{center}
\end{table}

\subsubsection{Cost and Delay Analysis of Command Propagation}
\label{sec:LNBotPropagationResults}

To assess the command propagation delay, we sent the following SYN flooding attack command to the C\&C servers from the botmaster (omitting the start and end characters):

\texttt{sudo hping3 -i u1 -S -p 80 -c 10 192.168.1.1}

We sent this command using both of the encoding methods we proposed earlier. For Huffman coding, we compared several different base number systems. The best result was obtained by using the Quaternary numeral system, the codebook of which is shown in Table \ref{tab:quaternaryhuffman}.

\begin{table}[h]
\vspace{-2mm}
  \begin{center}
    \caption{Obtained Codebook for Huffman Coding}
    \vspace{-3mm}
    \label{tab:quaternaryhuffman}
    \resizebox{0.77\linewidth}{!}{
    \begin{tabular}{|c c|c c|c c|c c|}
    \hline
    `s' & 234 & `n' & 233 & `o' & 232 & `h' & 231 \\ \hline 
    `d' & 224 & `g' & 223 & `c' & 222 & `9' & 221 \\ \hline 
    `6' & 214 & `2' & 213 & `3' & 212 & `u' & 211 \\ \hline 
    `p' & 144 & `i' & 143 & `8' & 142 & `0' & 141 \\ \hline 
    `.' & 24  & `1' & 12  & `-' & 13  & `E' & 4   \\ \hline 
    ` ' & 11  & `S' & 3   &     &     &     &     \\ \hline
    \end{tabular}
  }
  \vspace{-5mm}
  \end{center}
\end{table}

\begin{table}[h]
\vspace{-2mm}
  \begin{center}
    \caption{Breakdown of How Many Payments are Sent for the \\ \texttt{sudo hping3 -i u1 -S -p 80 -c 10 192.168.1.1} \\ Command with ASCII and Huffman Encoding}
    \vspace{-3mm}
    \label{tab:ASCII}
    \resizebox{0.9\linewidth}{!}{
    \begin{tabular}{|c|c|c|}
    \hline
    ~                & \multicolumn{2}{c|}{\textbf{Number of Payments}} \\ \cline{2-3}
    \textbf{Command} & \textbf{ASCII}        & \textbf{Quaternary}     \\ 
    ~                & \textbf{Encoding}     & \textbf{Huffman Encoding}     \\ \hline
    `sudo '          & 5             & 14     \\ \hline 
    `hping3 '        & 7             & 20     \\ \hline
    `-i '            & 3             & 7      \\ \hline
    `u1 '            & 3             & 7      \\ \hline
    `-S '            & 3             & 5      \\ \hline
    `-p '            & 3             & 7      \\ \hline
    `80 '            & 3             & 8      \\ \hline
    `-c '            & 3             & 7      \\ \hline
    `10 '            & 3             & 7      \\ \hline
     `192.168.1.1'   & 11            & 26     \\ \hline
    Total Number of  & 44            & 108    \\
    Payments         & ~             & ~      \\ \hline \hline
    Total Satoshi Used  & 2,813       & 215    \\ \hline
    \end{tabular}
  }
  \vspace{-2mm}
  \end{center}
\end{table}

\noindent \textbf{Cost Analysis:}
To calculate the cost of sending a command, we need to encode each character with its corresponding satoshi value. To give an example; the command \texttt{sudo} would be encoded as (115,117,100,111) with ASCII and (2,3,4,2,1,1,2,2,4,2,3,2) with Huffman where each value represents the satoshi amount of the payment. As can be seen in Table \ref{tab:ASCII}, to send the SYN flooding attack command, the botmaster spent 2,813 satoshi using the ASCII encoding and only 215 satoshi using the Huffman coding. Table \ref{tab:ASCII} shows how many payments were sent for each piece of the command as well as the total satoshi spent for sending the whole command for both encoding methods. While in both cases the botmaster will be reimbursed at the very end, we would like to note that the lifetime of the channels is closely related with these costs. In case of the ASCII encoding, the initial funds will be spent faster and the botmaster needs to reconfigure (or rebalance) the channels for continuous operation of the botnet. In case of the Huffman coding, this is not the case as the consumption of the channel funds is much slower. So, we can see that if channel lifetime is an important factor for the botmaster, the Huffman coding could be preferred. In other words, the Huffman coding gives the botmaster the ability to perform more attacks without creating high capacity channels.

However, the situation is reverse in case of routing fees. According to our experiments in LN, the average forwarding fee for payments was 4 satoshi. Taking this into account, the total forwarding fee for sending the SYN flooding attack command for different number of C\&C servers is shown in Table \ref{tab:routingfees}. The increase in the routing fees is linear for both the ASCII and Huffman coding. For 100 C\&C servers, total routing fee paid is only 17,600 satoshi for ASCII while it is 43,200 satoshi for the Huffman coding. This indicates that routing fees with Huffman is 2.45 times higher than ASCII. However, the satoshi spent for sending the actual command is 13 times less than ASCII. Thus, if the botmaster is willing to open channels with higher capacities, it is more feasible to run the botnet using the ASCII method because of the less routing fees paid.

\begin{table}[htb]
\vspace{-3mm}
  \begin{center}
    \caption{Routing Fees for Different Number of C\&C Servers}
    \vspace{-3mm}
    \label{tab:routingfees}
    \resizebox{0.84\linewidth}{!}{
    \begin{tabular}{|c|c|c|c|}
    \hline
      \makecell{\textbf{Number of} \\ \textbf{C\&C Servers}} & \makecell{\textbf{Routing Fees} \\ \textbf{(ASCII)}} & \makecell{\textbf{Routing Fees} \\ \textbf{(Huffman)}} \\
      \hline
      10 & 1,760 satoshi & 4,320 satoshi \\
      \hline
      25 & 4,400 satoshi & 10,800 satoshi \\
      \hline
      50 & 8,800 satoshi & 21,600 satoshi \\
      \hline
      100 & 17,600 satoshi & 43,200 satoshi \\
      \hline
    \end{tabular}
  }
  \vspace{-2mm}
  \end{center}
\end{table}

\begin{figure}[h]
\vspace{-3mm}
    \centering
    \includegraphics[width=\linewidth]{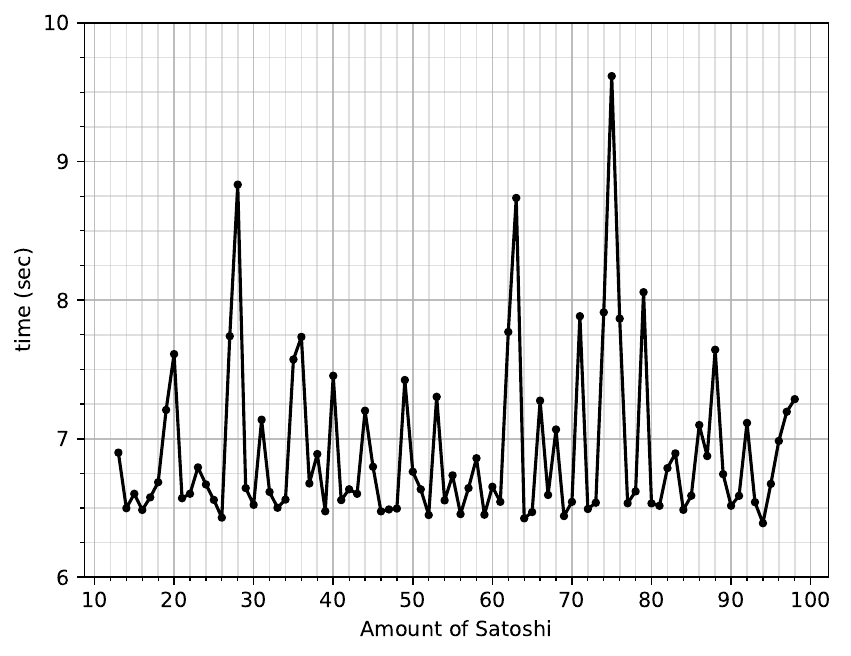}
    \vspace{-9mm}
    \caption{Time for \textit{key send} payments to reach their destinations with varying satoshi.}
    \label{fig:paymenttime}
\end{figure}

\noindent \textbf{Delay Analysis:} The propagation time of a command is calculated by multiplying the number of payments with the average delivery time of the payments. To estimate the average delivery time, we sent 90 \textit{key send payments} with different amounts from botmaster to our C\&C servers over LN at random times and measured the time it took for payments to reach their destinations. The results are depicted in Fig. \ref{fig:paymenttime}. 

As shown, key send payments took 7.156 seconds on average to reach their destinations and the maximum delay was never exceeding 10 seconds. This delay varies since it depends on the path being used and the load of each intermediary node in the LN. We observed that the number of hops for the payments was 4, which helps to strengthen unlinkability of payments and endpoints in case of any payment analysis in LN.

Using an average of 7.156 seconds, the total propagation time for the ASCII encoded payments is 7.156x44=314.864 seconds while it is 7.156x108=772.848 seconds for the Huffman coding. The Huffman coding reduces the Bitcoin spent for sending the commands (not the routing fees), but increases the command sending delays. 

This analysis is for sending the command to a single C\&C server only. If we generalize it to $n$ C\&C servers where the command comprises of $a$ characters; we can see that the time complexity of sending the command to all the C\&C servers is in the order of $O(na)$. We will call this \textit{total command sending delay} for later reference which basically stands for the time it takes for all the C\&C servers in the botnet to receive the command. Note that, we assume that the botmaster do not \textit{parallelize} the command sending by trying to send the command to multiple C\&C servers at the same time. Instead, s/he sends the command to each C\&C server sequentially.

Next, we analyze the iLNBot which uses the noise plugin for sending the commands instead of the ASCII or Huffman methods.

\subsection{iLNBot Analysis}
\label{sec:LNBot_noise}

Using the noise plugin for command sending dramatically reduces both the delay and cost of sending the commands. For the same SYN flooding attack command, this time the botmaster just sends a single payment to each C\&C server instead of sending consecutive payments for each character in the command. According to our experiments, the forwarding fee paid for a message sent using the noise plugin was 2 satoshi over an average of 30 tries. Thus, for 100 C\&C servers, the total forwarding fee paid for sending the same SYN flooding attack command is only 200 satoshi. In LNBot, it was 17,600 satoshi for ASCII and 43,200 satoshi for the Huffman method. Thus, there is an improvement around 98\% compared to the LNBot. 

The delay of sending the command decreases in similar proportions. According to our measurements in LN, the delay of sending messages using the noise plugin was only 2.11 seconds over an average of 30 tries. Length of the messages did not affect the delay. One thing to note here is that the delay is less compared to what we measured for the LNBot (7.156 seconds) even though both utilize the key send payments. This is due to the network conditions at different times as well as the number of hops used to send the payments. If we consider the ASCII encoding case where it took 314.864 seconds to send the SYN flooding attack command, the new method is 99\% faster than the ASCII method. Now, we can easily generalize the total command sending delay when there are $n$ C\&C servers in the botnet. Since the time complexity of sending the command to an individual C\&C server is $O(1)$, sending it to $n$ C\&C servers is in the order of $O(n)$. This is much faster than LNBot's $O(na)$ total command sending time.

\subsection{D-LNBot Evaluation}

Evaluation of D-LNBot is slightly different than the LNBot as we propose to utilize the Bitcoin Testnet. Because of this design choice, the cost just becomes zero. Thus, we only analyze the delay of command propagation. In this direction, we first analyze the time complexity of propagating a command to all the C\&C servers in D-LNBot. Then, we numerically show the total command propagation delay for varying number of C\&C servers based on the simulations we performed.

The average delay of sending a message using the noise plugin was already given at the previous section which was 2.11 seconds. In D-LNBot, this corresponds to the delay of sending a command from one C\&C server to another. In order to numerically calculate the total command propagation time of D-LNBot, we first need to analyze how the commands propagate among the C\&C servers in a D-LNBot topology.

D-LNBot's topology forms a connected graph where each vertex is a C\&C server and each edge is a logical neighbor link. To compute the time complexity of the command propagation in this graph, we first examine the two worst cases: 1) $m=1$ and 2) $m=n$ where $m$ is the number of active C\&C servers and $n$ is the number of C\&C servers. The topologies formed for these cases are shown in Fig. \ref{fig:D-LNBot-time-complexity} below.

\begin{figure}[h]
\vspace{-3mm}
    \centering
    \includegraphics[width=0.9\linewidth]{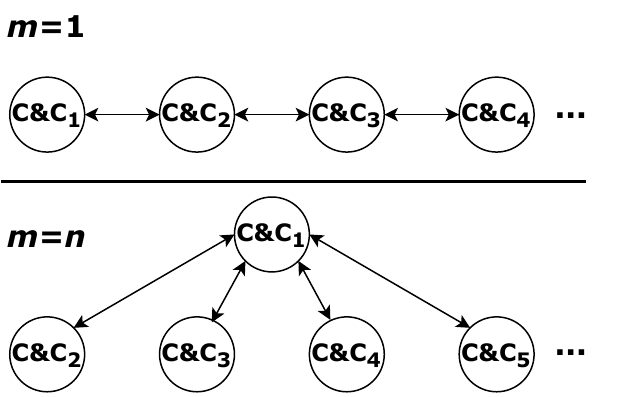}
    \vspace{-3mm}
    \caption{Logical topologies of the two worst cases in D-LNBot.}
    \label{fig:D-LNBot-time-complexity}
    \vspace{-1mm}
\end{figure}

In both of these cases, the time complexity of propagating a command to all the C\&C servers will be $O(n)$. In $m=1$ case, C\&C servers form a straight line and the total propagation time is directly proportional to the number of C\&C servers lined up as the command have to go through all the C\&C servers one by one until reaching the last C\&C server. Looking at the second topology where $m=n$, we see that it is the same centralized topology in LNBot and iLNBot. Thus, the time complexity of this case will also be $O(n)$ due to sequential propagation of the commands.

For $1<m<n$ case that is of the real interest to us, we already provided two example topologies at Fig. \ref{fig:commandProp}. At first glance, the topologies hint us that the time complexity of command propagation is at least in the order of $m$ (i.e., $\Omega(m)$). This is because we need at least $m$ messages in any case due to minimum number of neighbors we need to reach sequentially for a node. But as seen in Fig. \ref{fig:D-LNBot-time-complexity}, the height of the tree also has a role in the number of messages since we need to use a \textit{spanning tree} to reach every C\&C server. The height of the spanning tree grows as $m$ grows and it reduces as $m$ decreases. This corresponds to a height of $log_m n$. Now, the first level of the tree uses $m$ branches which already brings $m$ messages to the total complexity. Starting from level 2 of the tree, we get a height of $log_m n$ and each level adds $m$ more messages. This means we have $m(log_m n -1)$ more messages added to the complexity. Thus in total, our message complexity is $m + m(log_m n -1)$ which is in the order of $O(mlogn)$. This is faster than $O(n)$ in the average cases considering random topologies.  

We can conclude that, D-LNBot has a clear advantage over LNBot on both the cost and delay of sending the commands thanks to its unique distributed design and utilization of the Bitcoin Testnet.

\vspace{1mm}
\textbf{Simulation Results:} Measuring the command sending delay between two C\&C servers was straightforward however measuring the \textit{total} command sending delay is not as straightforward since it requires a rather complicated setup to perform the experiments. Specifically, one would need to create an extensive number of machines and program each of them to properly perform the botnet formation and the command propagation operations. Additionally, since in our design, the C\&C servers are created spontaneously, it is hard to realize this behavior on actual LN without actually releasing a malware in-the-wild. Instead of going this route, we performed simulations to measure the total command propagation delay in D-LNBot. Simulations also let us capture and test a wide range of cases. In this direction, we first simulated the D-LNBot formation phase in Python using the latest available LN topology at \cite{lngossip} which consists of 13,772 nodes and 118,021 channels.
In the simulation, all the required parameters such as number of active C\&C servers, number of C\&C servers, number of innocent nodes are set accordingly and a specific policy was set for C\&Cs to discover each other. After running the simulation, we observed that all the C\&C servers discovered the required number of neighboring servers and a D-LNBot was successfully formed. The resulting topologies can be reproduced by running the Python scripts in our GitHub page. As explained in Section \ref{sec:dlnbot_forming}, C\&C servers first found the most connected nodes (i.e., innocent nodes) in the network and randomly opened a channel to one of them. Additionally, each C\&C server randomly opened a channel to one of the nodes in the network to be able to route the payments even after closing its channel with the innocent node. This means that the C\&C servers did not necessarily establish channels to the centralized nodes in the network which actually resulted in most of the payments being forwarded in 5 to 7 hops. And in the simulations, we assumed that each hop introduces a delay between 0.4 and 0.5 seconds which is a realistic approximation from real LN payments we performed.

To be able to simulate the multi-hop mechanism, we first calculated the shortest path between a sender and the recipient using Dijkstra's algorithm \cite{dijkstra1959note}. Then, we created a data structure for the C\&C servers to be able to carry the hop information for the payments as well as a message. In this way, C\&C servers know where to forward the payment next and can deliver the message to the destination. The number of active C\&C servers in the simulations were 3. Finally, to be able to compare the results with LNBot and iLNBot, we varied the number of C\&C servers in the simulations. 

The results of total command propagation times are shown in Table \ref{tab:d-lnbot-time}. To compare, in LNBot, sending the command to a single C\&C server using the faster ASCII method took 314.864 seconds. Consequently, sending it to 10 C\&C servers takes 3,148.64 seconds. This is much slower than D-LNBot's 14 seconds. In iLNBot, sending the command to each C\&C server took 2.11 seconds. Thus, the total command sending delay for 10 C\&C servers is 21.1 seconds which is still slower than D-LNBot as expected. The results for more C\&C servers can be seen at Table \ref{tab:d-lnbot-time}. D-LNBot is faster than the other two due to its distributed topology that splits into branches. In a sense, command sending is parallelized with each branch enabling D-LNBot to achieve a faster propagation. Thus, we can conclude that D-LNBot propagates the commands to the C\&C servers much faster compared to LNBot and the gap between the two opens up even more as the number of C\&C servers increases.

\begin{table}[h]
  \begin{center}
    \caption{Total Command Propagation Time of SYN Flooding Attack Command in LNBot, iLNBot and D-LNBot for Different Number of C\&C Servers}
    \vspace{-3mm}
    \label{tab:d-lnbot-time}
    \resizebox{0.9\linewidth}{!}{
    \begin{tabular}{|c|c|c|c|c|c|}
    \hline
      \makecell{\textbf{Number of} \\ \textbf{C\&C Servers}} & \textbf{D-LNBot} & \textbf{iLNBot} & \textbf{LNBot}  \\
      \hline
      10 & 14 sec & 21.1 sec & 3,148.64 sec \\
      \hline
      25 & 36.5 sec & 52.75 sec & 7,871.6 sec \\
      \hline
      50 & 72.9 sec  & 105.5 sec & 15,743.2 sec \\
      \hline
      100 & 136.5 sec & 211 sec & 31,486.4 sec \\
      \hline
    \end{tabular}
  }
  \vspace{-6mm}
  \end{center}
\end{table}

\begin{table*}[h]
  \begin{center}
    \caption{Comparison of LNBot, iLNBot and D-LNBot with Similar Bitcoin-based Botnets}
    \vspace{-3mm}
    \label{tab:comparison}
    \resizebox{0.9\linewidth}{!}{
    \begin{tabular}{|c|c|c|c|c|c|}
    \hline
      \textbf{Botnet} & \textbf{Network} & \textbf{Method} & \textbf{Cost} & \textbf{Delay} & \textbf{Scalability}\\
      \hline
      \textbf{BOTCHAIN \cite{Botchain}} & Mainnet & OP\_RETURN & \makecell{22,848 satoshi \\ as of today} & $\sim$ 1 hour & Low \\
      \hline
      \textbf{ZombieCoin 2.0 \cite{ali2018zombiecoin}} & Mainnet & \makecell{OP\_RETURN +\\ subliminal channels} & \makecell{22,848 satoshi \\ as of today} & $\sim$ 10 seconds & Low \\
      \hline
      \textbf{DUSTBot \cite{zhong2019dustbot}} & \makecell{Both Mainnet \\ and Testnet} & OP\_RETURN & \makecell{22,848 satoshi \\ as of today} & $\sim$ 10 seconds & Medium \\
      \hline
      \textbf{ChainChannels \cite{frkat2018chainchannels}} & Mainnet & ECDSA Signature & \makecell{$>$22,848 satoshi \\ as of today} & $\sim$ 10 seconds & Medium \\
      \hline
      \textbf{CoinBot \cite{yin2020coinbot}} & \makecell{Either Mainnet \\ or Testnet} & \makecell{OP\_RETURN +\\ website} & 226 satoshi & $\sim$ 6 minutes & Low \\ 
      \hline
      \textbf{Franzoni et al. \cite{franzoni2020leveraging}} & Testnet & \makecell{non-standard \\ OP\_RETURN} & \makecell{133-51,349 \\ Testnet satoshi} & $\sim$ 10 minutes & Low \\
      \hline
      \textbf{LNBot} & Mainnet & LN Payment & \makecell{176 satoshi \\ (per C\&C)} & \makecell{$\sim$ 5 minutes \\ (per C\&C)} & High \\
      \hline
      \textbf{iLNBot} & Mainnet & LN Payment & \makecell{2 satoshi \\ (per C\&C)} & \makecell{$\sim$ 2 seconds \\ (per C\&C)} & High \\
      \hline
      \textbf{D-LNBot} & Testnet & LN Payment & 2 Testnet satoshi & \makecell{$\sim$ 1.5 seconds \\ (per C\&C)} & High \\
      \hline
    \end{tabular}
  }
  \end{center}
  \vspace{-5mm}
\end{table*}

\subsection{Comparison with Other Similar Botnets}

We also considered other existing botnets that utilize Bitcoin for their command and control. The extensive comparison of these botnet with our proposed LNBot, iLNBot and D-LNBot are presented in Table \ref{tab:comparison}. We considered the following metrics for the comparison: 1) Which Bitcoin network is utilized by the botnet?; 2) What methods are used for command propagation?; 3) What is the cost of sending our SYN flood attack command to all the bots in the botnet?; 4) How long does it take for all the bots in the botnet to receive the SYN flood attack command from the botmaster (i.e., total command propagation time); 5) How scalable is the botnet? Low: thousands of bots, Medium: hundreds of thousands of bots, High: millions of bots.

Before interpreting the results, it is important to note that the cost of sending the commands is variable for some botnets because they utilize on-chain Bitcoin transactions whose fees change dynamically depending on the load on the Bitcoin network. At the time of writing this paper, fee for a Bitcoin transaction with a median transaction size of 224 bytes was 22,848 satoshi\footnote{\url{https://bitcoinfees.earn.com/}}. For the botnets that fully run on Bitcoin's Testnet, the cost of command sending is normally zero but in order to have a quantitative comparison with the other botnets, we still mentioned the cost with \textit{Testnet satoshi}.

As can be seen in Table \ref{tab:comparison}, BOTCHAIN \cite{Botchain} is the worst among all with a high cost and long delay for sending the commands. ZombieCoin 2.0 \cite{ali2018zombiecoin}, DUSTBot \cite{zhong2019dustbot} and ChainChannels \cite{frkat2018chainchannels} are faster but still come with high costs. CoinBot \cite{yin2020coinbot} has much less cost due to using 1 sat/byte fee for Bitcoin transactions however, it still takes a long time to propagate the commands. Franzoni et al. \cite{franzoni2020leveraging} can send the commands for free due to using Testnet however still has long delays. LNBot spends less than all previously mentioned botnets but still has long delays. iLNBot achieves much faster and cheaper command sending compared to LNBot due to the use of noise plugin. Finally, D-LNBot sends the commands even faster than iLNBot and for free ultimately being the best among all. 

If we look at the scalability of these botnets, only DUSTBot \cite{zhong2019dustbot} and ChainChannels \cite{frkat2018chainchannels} seem to be able to somewhat scale due to the way they propagate the commands to the bots. Others rely on blockchain confirmations of their Bitcoin transactions which greatly limits their scalability. Our proposed LNBot, iLNBot and D-LNBot on the other hand can scale freely thanks to their 2-layer architecture.

\section{Security \& Anonymity Analysis and Countermeasures}
\label{sec:countermeasures} 

In this section, we discuss the security properties of LNBot and D-LNBot as well as possible countermeasures to detect their activities in order to minimize their impacts.

\vspace{1mm}
\noindent $\bullet$ \textit{Taking LN Down:} Obviously, the simplest way to eliminate LNBot and D-LNBot is to take down LN as a whole once there is any suspicion about a botnet. However, this is very unlikely due to LN being a very resilient decentralized payment channel network. In addition, today many applications are running on LN and shutting it down may cause a lot of financial loss for numerous stakeholders.

\vspace{1mm}
\noindent $\bullet$ \textit{Resetting the Bitcoin Testnet:} Since taking LN down is not really an option, taking down the Bitcoin Network itself could be explored. Since its creation, Bitcoin Mainnet never went offline. Bitcoin Testnet on the other hand is in its third iteration, which is called Testnet3. Previous two Testnets were reset due to some technical reasons\footnote{\url{https://bitcoin.stackexchange.com/questions/36252/testnet-version-history}}. Essentially, this poses a great risk for a botnet running on the Bitcoin Testnet. When the existence of D-LNBot is discovered, it is possible that the Bitcoin developers might go the route of resetting the Testnet again. Basically, a new genesis block is generated invalidating all the previous blocks. If this happens, all LN channels created on the Bitcoin Testnet will become unusable. This action therefore will stop the D-LNBot completely.

\vspace{1mm}
\noindent $\bullet$ \textit{Compromising and Shutting Down a C\&C Server:} In our design, there are many C\&C servers each of which is controlling a mini-botnet. Given the past experience with various traditional botnets, it is highly likely that these mini-botnets will be detected at some point in the future paving the way for also the detection of a C\&C server. This will then result in the revelation of its location/IP address and eventually physical seizure of the machine by law enforcement. Nevertheless, in LNBot, the seizure of a C\&C server will neither reveal the identity of the botmaster nor other C\&C servers since a C\&C server receives the commands through onion routed payments catered with Sphinx's secure packet format, which does not reveal the original sender of the payments. Additionally, the communication between the botmaster and the C\&C servers is 1-way meaning that botmaster can talk to the C\&C servers, but servers cannot talk back since they do not know the LN address of the botmaster. This 1-way communication combined with the onion routed payments ensure that the identity of the botmaster will be kept secret at all times.

Note that, in LNBot, since the C\&C servers hold the LN public key of the collector, it will also be revealed when a C\&C server is compromised. However, since the collector's LN channels are all private, its IP address or location is not known by the C\&C servers. Therefore, learning the LN public key of the collector node does not help locating it physically. However, honest LN nodes in the network can be informed of collector's LN public key and be advised to refuse opening channels to it. In this case, collector might be unable to receive funds from the C\&C servers. Therefore, we offer alternatives to address this issue with the new features of LN. Best and most reliable option is to use \textit{splicing} which enables either adding or removing funds from a channel by a single on-chain transaction without having to close the channel or create a new one. It is one of the most anticipated additions to LN because it will solve the biggest problem of LN which is the channel liquidity management. In case of C\&C servers, they would \textit{splice out} their channel funds to an on-chain output which the botmaster can spend and re-use. Currently there is an experimental implementation of splicing\footnote{\url{https://github.com/ElementsProject/lightning/pull/5675}}. Thus, we rather recommend using \textit{PeerSwap} protocol instead until splicing is fully integrated into LN. PeerSwap enables swapping on-chain Bitcoins for inbound or outbound channel liquidity between a channel's peers \cite{peerswap}. The protocol is completely trustless, low cost and can be used today with both \textit{Core Lightning} and \textit{lnd}. C\&C servers can \textit{swap out} their channels funds for on-chain Bitcoins which the botmaster can spend and re-use. Eventually, we can see that taking down a single C\&C server shuts down the LNBot partially resulting in less damage to victims.

In case of the D-LNBot, revealing a C\&C server will also reveal its neighboring C\&C servers' LN public keys and possibly IP addresses. Even though more information is being revealed, it is easier to recruit new C\&C servers in D-LNBot. The reason for that is, in D-LNBot, botmaster does not create the C\&C servers himself rather they are created through \textit{malware\_1} infections. To conclude, we can say that, taking down a single C\&C server only reveals a few other C\&C servers which is not critical considering that the number of C\&C servers in D-LNBot will likely to be greater than that of LNBot. Regardless, this still takes down a portion of D-LNBot which helps reducing its impacts on the victims.

Additionally, in D-LNBot, when a C\&C server is compromised, the function $f()$ and the value $\xi$ can be revealed. In such a case, the active C\&C servers' LN public keys (and possibly IP addresses) will be revealed. As explained in Section \ref{sec:dlnbot_forming}, there are only \textit{m} active C\&C servers in D-LNBot at any time, thus revelation of these hardcoded parameters will only cause revelation of \textit{m} C\&C servers. Older C\&C servers that already closed their channels with the innocent nodes cannot be detected with this information. Thus, an observer could instead start monitoring the new channel openings in LN to identify the newly joined C\&Cs. However, s/he cannot know for certain if an LN node satisfying the policy is a C\&C server or a honest LN node. Because, the channel capacities satisfying the policy can also be used by honest LN nodes. Thus, we can conclude that revelation of these parameters expose active C\&C servers helping investigators take down a portion of D-LNBot.

\vspace{1mm}
\noindent $\bullet$ \textit{Payment Flow Timing Analysis for Detecting the Botmaster:} 
Our original analysis for this attack in \cite{kurt2020lnbot} is applicable to D-LNBot as well. A similar topology can be created around a captured C\&C server in an attempt to reveal the sender of the payments. However, the senders are already known by the C\&C server; i.e., the neighboring C\&C servers. So, doing this timing analysis to try to find the botmaster will be pointless as s/he is never involved in the process. Only case where this might be useful is applying the analysis to the C\&C server that initially receives the command from the botmaster. However, even in that case, C\&C server will only receive a single payment for each command which will make the timing analysis impracticable as the analysis requires the C\&C server to receive a number of payments to possibly provide useful information.

\vspace{1mm}
\noindent $\bullet$ \textit{Poisoning Attack:} 
In LNBot, another effective way to attack the botnet is through message poisoning. Basically, once a C\&C server is compromised, its LN public key will be known. Using the public key, one can send payments to the C\&C server at the right time to corrupt the messages sent by the botmaster. There is currently no authentication mechanism that can be used by the botmaster to prevent this issue without being exposed. Recall that the commands are encoded into a series of payments and when an attacker sends a different character during a command transmission, it will corrupt the syntax and thus eventually the command will not have any effect on the corresponding mini-botnet. The right time will be decided by watching the payments and packets arriving at the C\&C server. The disadvantage of this, however, is that one needs to pay for those payments. Nonetheless, this can be an effective way to continue engaging with the botmaster for detection purposes rather than just shutting down the C\&C server while rendering any attack impossible. 

iLNBot and D-LNBot are not vulnerable to this attack because of the authentication mechanism that exists in the noise messages as explained in Section \ref{sec:dlnbotcommandpropagation}. Even if somehow an attacker gets the LN public keys of the C\&C servers and initiates commands to them using the noise plugin, the C\&C servers will realize that the commands are coming from an unknown node so, they can just ignore them. Additionally, an attacker cannot poison the commands as the commands are sent with a single payment. Rather, the attacker can only try to execute his/her own commands on the C\&C servers which will not be possible due to the authentication mechanism.

\vspace{1mm}
\noindent $\bullet$ \textit{Analysis of On-chain Transactions:} Another possible attack vector to both LNBot and D-LNBot is to analyze the on-chain Bitcoin transactions of the C\&C servers (i.e., channel opening and closing transactions). For such forensic analysis, the Bitcoin addresses of the C\&C servers have to be known. As with many other real-life botnets, botmasters generally use Bitcoin mixers to hide the source of the Bitcoins. Usage of such mixers makes it very challenging to follow the real source of the Bitcoins since the transactions are mixed between the users using the mixer service \cite{van2018bitcoin}. Even though, the chances of finding the identity of the botmaster through this analysis is low, it can provide some useful information to law enforcement.

\vspace{1mm}
\noindent $\bullet$ \textit{Packet Inspection:} Finally, LN payments could also be analyzed with packet inspection tools such as Wireshark. However, this will prove to be not very useful because all LN payments are encapsulated into onion packets as explained in Section \ref{sec:onionrouting}. Any information that might be included in the payments such as messages, commands or instructions cannot be decrypted without having access to the associated private keys.

\section{Conclusion}
\label{sec:conclusion}
LN has been formed as a new payment network to address the drawbacks of Bitcoin transactions in terms of time and cost. In addition to the fast and cheap transactions, LN provides a perfect opportunity for covert communications as no transactions are recorded in the blockchain. Therefore, in this paper, we proposed a new covert hybrid botnet by utilizing the LN payment network formed for Bitcoin operations. The idea was to control the C\&C servers through commands that are sent in the form of LN payments. The proof-of-concept implementation of this architecture indicated that LNBot can be successfully created and commands for attacks can be sent to C\&C servers through LN with negligible costs yet with very high anonymity. To improve on the cost and delay of sending the commands, we proposed a slightly modified version of LNBot which utilizes a new feature of LN that enables sending messages by embedding them into payments. Our evaluation results showed an 98\% improvement in both cost and delay metrics. Finally, we proposed D-LNBot, a cost-free and distributed version of LNBot with faster command sending times. Our evaluation results showed that, commands are propagated to all the C\&C servers in $O(mlogn)$ time compared to LNBot's $O(na)$. To minimize the impact of these proposed botnets, we offered several countermeasures that include the possibility of searching for the botmaster.

\appendices


\ifCLASSOPTIONcompsoc
  \section*{Acknowledgments}
\else
  \section*{Acknowledgment}
\fi

We would like to thank Christian Decker for helping us better understand various aspects of LN.

\ifCLASSOPTIONcaptionsoff
  \newpage
\fi





\bibliographystyle{IEEEtran}
\bibliography{IEEEabrv,arxiv_v2_references}

%



%
\begin{IEEEbiography}[{\includegraphics[width=1in,height=1.25in,clip,keepaspectratio]{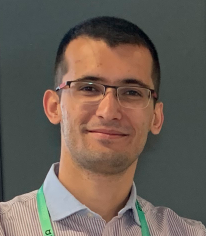}}]{Ahmet Kurt} received two B.S. degrees from Antalya Bilim University, Antalya, Turkey in 2018. He received the M.Sc. degree from the Department of Electrical and Computer Engineering, Florida International University, Miami, Florida in 2021. He is currently pursuing the Ph.D. degree in Electrical and Computer Engineering with the Florida International University, Miami, United States. His current research interests include Bitcoin’s lightning network, Bitcoin and wireless networks.
\end{IEEEbiography}


\begin{IEEEbiography}[{\includegraphics[width=1in,height=1.25in,clip,keepaspectratio]{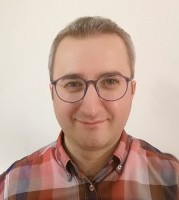}}]{Enes Erdin} is an Assistant Professor in the Computer Science Department at University of Central Arkansas, Conway. He conducts research in the areas of hardware security, blockchain technology, and cyber-physical systems. Erdin received a Ph.D. in Electrical and Computer Engineering from Florida International University, Miami where he was a NSF CyberCorps fellow.
\end{IEEEbiography}


\begin{IEEEbiography}[{\includegraphics[width=1in,height=1.25in,clip,keepaspectratio]{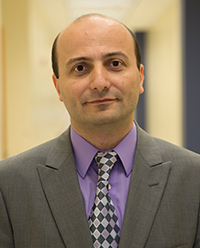}}]{Kemal Akkaya} received the Ph.D. degree in computer science from the University of Maryland, Baltimore, MD, USA, in 2005. He joined, as an Assistant Professor, the Department of Computer Science, Southern Illinois University Carbondale (SIU), Carbondale, IL, USA, where he was an Associate Professor from 2011 to 2014. He was also a Visiting Professor with George Washington University, Washington, DC, USA, in 2013. He is currently a Professor with the Department of Electrical and Computer Engineering, Florida International University, Miami, FL, USA. His current research interests include security and privacy, IoT and cyber-physical systems. He was the recipient of the Top Cited Article Award from Elsevier in 2010. He is currently an Area Editor for the Elsevier Ad Hoc Network journal, and is on the Editorial Board of the IEEE Communication surveys and tutorials. 
\end{IEEEbiography}

\begin{IEEEbiography}[{\includegraphics[width=1in,height=1.25in,clip,keepaspectratio]{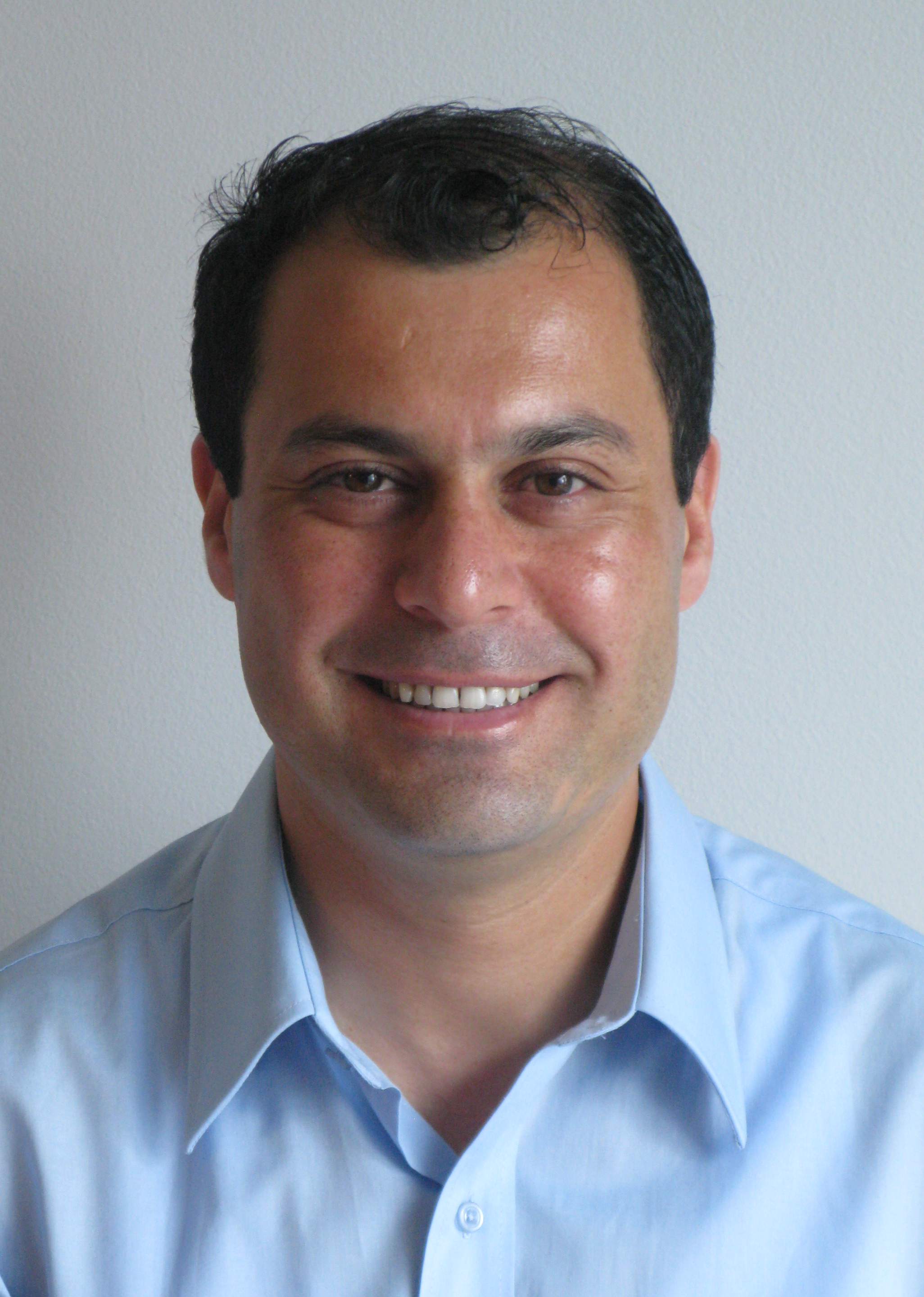}}]{A. Selcuk Uluagac} is an Eminent Scholar Chaired Associate Professor and the director of Cyber-Physical Systems Security Lab in the Department of Electrical and Computer Engineering at Florida International University, Miami, Florida, USA. Before FIU, he was a Senior Research Engineer at Georgia Institute of Technology and at Symantec. He holds a M.S. and Ph.D. from Georgia Tech and an M.S. from Carnegie Mellon University. He is expert on security and privacy topics with hundreds of scientific/creative works in practical and applied aspects of these areas. He received US NSF CAREER Award (2015), US Air Force Office of Sponsored Research’s Summer Faculty Fellowship (2015), and University of Padova’s (Italy) Summer Faculty Fellowship (2016). His research in cybersecurity has been funded by numerous government agencies and industry. He has served on the program committees of top-tier security conferences such as IEEE Security \& Privacy (“Oakland”), NDSS, Usenix Security, inter alia. He was the General Chair of ACM Conference on Security and Privacy in Wireless and Mobile Networks (ACM WiSec) in 2019. Currently, he serves on the editorial boards of IEEE Transactions on Mobile Computing, Elsevier Computer Networks Journal, and the IEEE Communications and Surveys and Tutorials (network security lead).
\end{IEEEbiography}

\begin{IEEEbiography}[{\includegraphics[width=1in,height=1.25in,clip,keepaspectratio]{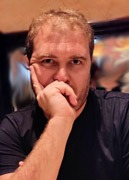}}]{Mumin Cebe} is an Assistant Professor in the Department of Computer Science at Marquette University. He has a Ph.D. degree from the Department of Electrical and Computer Engineering at Florida International University, US, and an M.Sc. from the Department of Computer Science at Bilkent University, Turkey.  Currently, he conducts research in security and privacy areas related to Internet-of-Things (IoT), blockchain, and cyber-physical Systems (CPS).
\end{IEEEbiography}


\vfill


\end{document}